\documentclass[
 reprint,
 amsmath,amssymb,
 aip,
 apl,
]{revtex4-1}

\usepackage{graphicx}
\usepackage{dcolumn}
\usepackage{bm}
\usepackage[utf8]{inputenc}
\usepackage[T1]{fontenc}
\usepackage{mathptmx}
\usepackage{etoolbox}
\usepackage[table]{xcolor}
\makeatletter
\def\@email#1#2{%
 \endgroup
 \patchcmd{\titleblock@produce}
 {\frontmatter@RRAPformat}
 {\frontmatter@RRAPformat{\produce@RRAP{*#1\href{mailto:#2}{#2}}}\frontmatter@RRAPformat}
 {}{}
}%
\makeatother
\begin{document}

\newcommand{\smc}{\mathrm{g}_{\uparrow \downarrow}}

\newcommand{\alphaeff}{\alpha_{\rm eff}}

\newcommand{\Hk}{H_{\rm k}}

\newcommand{\Meff}{M_{\rm eff}}

\newcommand{\Ms}{M_{\rm \textrm{s}}}

\newcommand{\tFM}{t_{\rm FM}}

\newcommand{\tHM}{t_{\rm HM}}

\newcommand{\tMnSn}{t_{\rm Mn_3Sn}}

\newcommand{\Ho}{\Delta H_{\rm 0}}

\newcommand{\Ks}{K_{\rm s}}

\newcommand{\SHC}{\sigma_{\rm SH}}

\newcommand{\sdl}{\lambda_{\rm Mn_3Sn}}

\newcommand{\Vdc}{V_{\rm dc}}

\newcommand{\VISHE}{V_{\rm ISHE}}

\newcommand{\Vsym}{V_{\rm sym}}

\newcommand{\Vasym}{V_{\rm asym}}

\newcommand{\DelH}{\Delta H}

\newcommand{\Hr}{H_{\rm \textrm{r}}}

\newcommand{\HR}{H_{\rm \textrm{R}}}

\newcommand{\jc}{\vec{J}_{\rm \textrm{C}}}

\newcommand{\js}{\vec{J}_{\rm S}}

\newcommand{\Lr}{\lambda_{\textrm{IREE}}}

\newcommand{\alphaR}{\alpha_{\textrm{R}}}

\newcommand{\TPSH}{\theta_\textrm{PSH}}

\newcommand{\SHA}{\theta_\textrm{SH}}

\newcommand{\IC}{I_{\rm C}}

\preprint{AIP/123-QED}

\title{Interfacial origin of unconventional spin-orbit torque in Py/$\gamma-$IrMn$_{3}$}

\author{Akash Kumar}
\thanks{These authors contributed equally.}
\affiliation{Department of Physics, Indian Institute of Technology Delhi, Hauz Khas, New Delhi 110016, India.}
\affiliation{Department of Physics, University of Gothenburg, Gothenburg,412 96, Sweden}

\author{Pankhuri Gupta}
\thanks{These authors contributed equally.}
\affiliation{Department of Physics, Indian Institute of Technology Delhi, Hauz Khas, New Delhi 110016, India.}

\author{Niru Chowdhury,}
\affiliation{Department of Physics, Indian Institute of Technology Delhi, Hauz Khas, New Delhi 110016, India.}
\author{Kacho Imtiyaz Ali Khan,}
\affiliation{Department of Physics, Indian Institute of Technology Delhi, Hauz Khas, New Delhi 110016, India.}
\author{Utkarsh Shashank,}
\affiliation{Department of Physics and Information Technology, Kyushu Institute of Technology, Iizuka, Fukuoka 820-8502, Japan.}
\author{Surbhi Gupta,}
\affiliation{Department of Physics and Information Technology, Kyushu Institute of Technology, Iizuka, Fukuoka 820-8502, Japan.}
\author{Yasuhiro Fukuma,}
\affiliation{Department of Physics and Information Technology, Kyushu Institute of Technology, Iizuka, Fukuoka 820-8502, Japan.}
\author{Sujeet Chaudhary}
\affiliation{Department of Physics, Indian Institute of Technology Delhi, Hauz Khas, New Delhi 110016, India.}
\author{Pranaba Kishor Muduli*}
 \email{muduli@physics.iitd.ac.in}
\affiliation{Department of Physics, Indian Institute of Technology Delhi, Hauz Khas, New Delhi 110016, India.}

\date{\today}

\begin{abstract}

Angle-resolved spin-torque ferromagnetic resonance measurements are carried out in heterostructures consisting of Py (Ni$_{81}$Fe$_{19}$) and a noncollinear antiferromagnetic quantum material $\gamma-$IrMn$_{3}$. The structural characterization reveals that $\gamma-$IrMn$_{3}$ is polycrystalline in nature. A large exchange bias of 158~Oe is found in Py/$\gamma-$IrMn$_{3}$ at room temperature, while $\gamma-$IrMn$_{3}$/Py and Py/Cu/$\gamma-$IrMn$_{3}$ exhibited no exchange bias. Regardless of the exchange bias and stacking sequence, we observe a substantial unconventional out-of-plane anti-damping torque when $\gamma-$IrMn$_{3}$ is in direct contact with Py. The magnitude of the out-of-plane spin-orbit torque efficiency is found to be twice as large as the in-plane spin-orbit torque efficiency. The unconventional spin-orbit torque vanishes when a Cu spacer is introduced between Py and $\gamma-$IrMn$_{3}$, indicating that the unconventional spin-orbit torque in this system originates at the interface. These findings are important for realizing efficient antiferromagnet-based spintronic devices via interfacial engineering.

\end{abstract}
\maketitle


Spin-orbit torques (SOTs) in ferromagnetic/heavy metal (FM/HM) heterostructures have emerged as a powerful tool for a wide range of spintronic applications.~\cite{Ando2008, gambardella2011current, Liu2012, VEDemidov2012, Emori2013, awad2016natphys} Large anti-damping SOTs are of utmost interest and play a pivotal role in ultrafast magnetization switching~\cite{Liu2012,garello2014ultrafast}, spintronic oscillators~\cite{VEDemidov2012,awad2016natphys,chen2016ieeeproc,behera2022energy,kumar2023robust} as well as the emerging area of spintronic-based neuromorphic computing.~\cite{zahedinejad2019two,houshang2020spin, garg2021kuramoto, kumar2022nanoscale,yadav2023demonstration} In the FM/HM system, the SOT is generated either via the spin Hall effect~\cite{Dyakonov1971, Hirsch1999, sinova2015Rev, Manchon2019review} from the HM layer and/or the Rashba-Edelstein effect~\cite{bychkov1984properties,edelstein1990spin} from the interface between the FM and HM layers. In both cases, when a charge current flows in the $x-$direction, it generates a spin current in the $z-$direction with the spin polarization along the $y-$direction, which can apply an in-plane anti-damping torque on the adjacent FM layer.~\cite{gambardella2011current, Manchon2019review} This is often referred to as  ``conventional SOT." Due to the symmetry of the conventional SOT, it is difficult to achieve deterministic switching of a perpendicularly magnetized system used in high-density magnetic recording. 
Recently, an out-of-plane damping-like torque was reported in WTe$_{2}$/Py by using the low crystalline symmetry of WTe$_{2}$.~\cite{Macneil2017NP_WTe2} Field-free switching in perpendicularly magnetized FMs using the unconventional spin-orbit torques (USOTs) in WTe$_{2}$ was also demonstrated recently.~\cite{xie2021field} 
Since then, USOTs have been reported in non-collinear antiferromagnets such as Mn$_{3}$GaN~\cite{nan2020controlling}, Mn$_{3}$SnN~\cite{you2021cluster}, Mn$_{3}$Sn~\cite{kondou2021giant} and IrMn$_{3}$~\cite{zhou2020magnetic} by using their low magnetic symmetry, which gives rise to an out-of-plane spin polarization for in-plane charge currents. The non-collinear antiferromagnetic also exhibits a large anomalous Hall effect due to the non-zero Berry curvature caused by the spin texture in the Kagome plane.~\cite{shindou2001orbital, kubler2014non, chen2014anomalous, nagaosa2010anomalous, higo2018anomalous, han2018quantum} In addition, the ferromagnetic/antiferromagnetic (FM/AFM) bilayer systems are also extensively explored for the presence of exchange bias~\cite{nogues1999exchange} and large SOTs, which lead to field-free switching of ferromagnets.~\cite{oh2016field,fukami2016magnetization,van2016field}
Particularly, for the non-collinear antiferromagnet IrMn$_3$, an unconventional torque is reported for epitaxial L1$_{\rm 0}-$IrMn and L1$_{2}-$IrMn$_{3}$.~\cite{zhou2020magnetic}

In this work, we investigate SOT in Py(= Ni$_{81}$Fe$_{19}$)/$\gamma-$IrMn$_{3}$, $\gamma-$IrMn$_{3}$/Py and Py/Cu/$\gamma-$IrMn$_{3}$ using the spin-torque ferromagnetic resonance (STFMR) technique and determine $\xi_{\rm DL}^{x}$, $\xi_{\rm DL}^{y}$ and $\xi_{\rm DL}^{z}$ which represents damping-like spin torque efficiency in $x-$, $y-$ and $z-$ direction, respectively. All the three components $\xi_{\rm DL}^{x}$, $\xi_{\rm DL}^{y}$, and $\xi_{\rm DL}^{z}$ are found to be present in Py/$\gamma-$IrMn$_{3}$, $\gamma-$IrMn$_{3}$/Py system, while only $\xi_{\rm DL}^{y}$ is found to be present in Py/Cu/$\gamma-$IrMn$_{3}$. This means that while USOT exists in both Py/$\gamma-$IrMn$_{3}$ and $\gamma-$IrMn$_{3}$/Py systems, the Py/Cu/$\gamma-$IrMn$_{3}$ system has only conventional SOT. Thus, the USOT in our case arises from the interface between Py and polycrystalline $\gamma-$IrMn$_{3}$, which is in stark contrast with previous findings where the USOT was observed only due to the presence of low magnetic symmetry in the bulk of IrMn$_3$.~\cite{zhou2020magnetic}  Our results show that USOT can be obtained using a CMOS-compatible naturally oxidized high resistive Si substrate, unlike previous studies of Zhou \textit{et al.},~\cite{zhou2020magnetic} for which the authors used specialized substrates, which is not compatible with CMOS-technology.

\begin{figure} [t!]
\centering
\includegraphics[width=\linewidth]{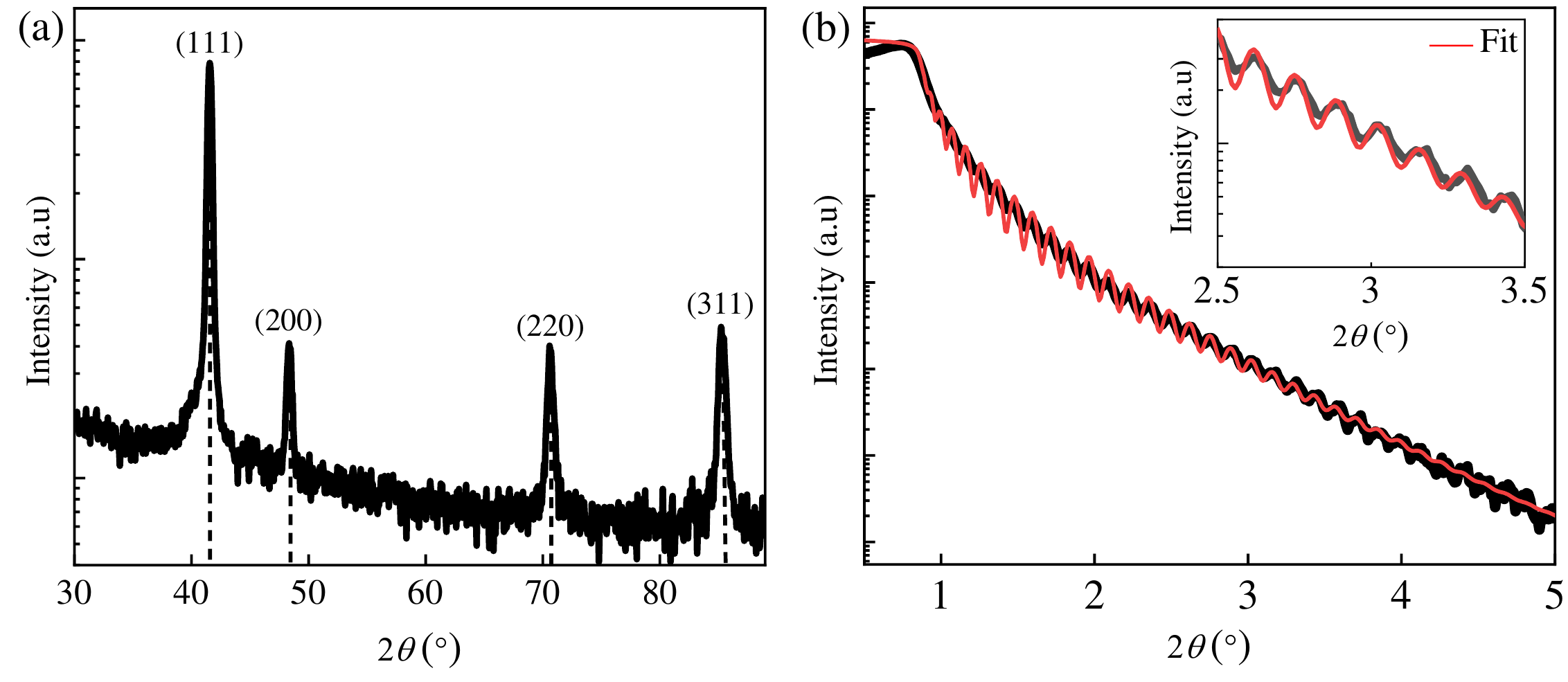}
\caption{ (a) Grazing-incidence X-ray diffraction spectrum measured in $\theta-2\theta$ mode, and (b) X-ray reflectivity spectra (black line) as well as the fit (red line) for a 62~nm thick IrMn$_3$ thin film. The inset in (b) shows a zoomed portion of XRR spectra with the fit (red line).}
\label{fig:1}
\end{figure}

The thin film heterostructures are grown using an AJA Orion 8 UHV magnetron sputtering system at room temperature. The base pressure of the sputtering system is better than $2\times10^{-8}$ Torr. Samples were grown at a working pressure of $1\times10^{-3}$ Torr. First, we optimized the growth of IrMn$_3$. We used a target with the composition of 20 at\% Ir and 80 at\% Mn. Subsequently, we deposited multilayer samples with Py and $\gamma-$IrMn$_{3}$ on Si substrates at room temperature. The structural characterization was performed using X-ray diffraction (XRD) using a PANalytical x-ray diffractometer equipped with Cu K$_{\alpha}$ ($\lambda$ = 1.5406 \AA) source. We use in-plane grazing incidence X-ray diffraction at an incidence angle of $1^{\circ}$ to determine the crystalline phase of IrMn, as shown in Fig.~\ref{fig:1}(a). The Bragg peaks are observed at $2\theta$ = 41.6$^{\circ}$, 48.3$^{\circ}$, 70.7$^{\circ}$, and 85.4$^{\circ}$, which correspond to the IrMn$_{3}$ (111), (200), and (220) and (311)-planes, respectively. The superlattice reflection of L1$_2$-IrMn$_3$, namely peaks corresponding to (110) and (211), are absent in our sample, indicating polycrystalline growth of only $\gamma-$IrMn$_3$ phase.~\cite{migliorini2018spontaneous,tsunoda2006l12} The thickness, density, and roughness of the films were analyzed using X-ray reflectivity (XRR) technique and fitted using recursive Parratt's formalism~\cite{parratt1954surface} as shown in Fig.~\ref{fig:1}(b). The thickness and roughness of $\gamma-$IrMn$_3$ in this thin film were found to be $\approx $ 62 nm and $<$ 0.3~nm, respectively. The optimal growth rate of $\gamma-$IrMn$_3$ was $\approx$ 0.9 \AA/s. The zoomed XRR spectrum and respective fit are shown in the inset of Fig.~\ref{fig:1}(b). From the XRR measurements, the interface roughness of multilayer structures was found to be $<$ 0.3~nm (not shown). To induce exchange bias, these structures were grown in the presence of a 300~Oe \textit{in-situ} in-plane magnetic field. To avoid oxidation, samples were capped with 2 nm Ta or Pt thin films. Microstrip devices of dimension ($4~\mu\rm m\times12~\mu\rm m$) were fabricated for the STFMR measurements using optical lithography followed by Ar-ion etching. Ground-signal-ground electrodes with 150 $\mu$m pitch are fabricated for contact pads. In the present manuscript, we discuss the results for three samples, denoted by Py/$\gamma-$IrMn$_{3}$: Py~(8.7~nm)/$\gamma-$IrMn$_{3}$(13~nm), $\gamma-$IrMn$_{3}$/Py: $\gamma-$IrMn$_{3}$(15 nm)/Py(20 nm), and Py/Cu/$\gamma-$IrMn$_{3}$: Py(10 nm)/Cu(1.5 nm)/$\gamma-$IrMn$_{3}$(13 nm).

\begin{figure*}[t!]
\centering
\includegraphics[width=\linewidth]{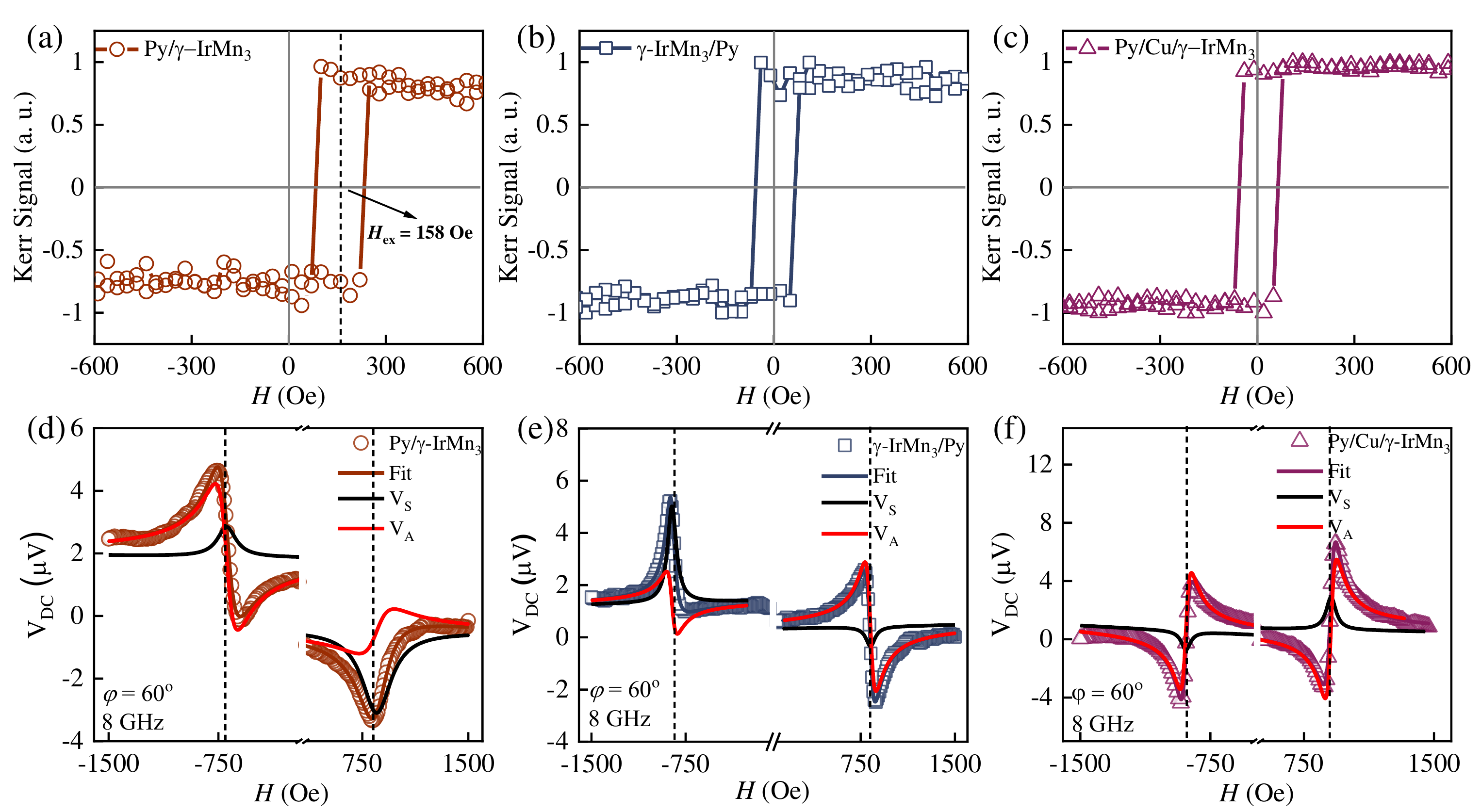}
\caption{Magnetic hysteresis loops measured using magneto-optic Kerr effect magnetometer for (a) Py/$\gamma-$IrMn$_{3}$, (b) $\gamma-$IrMn$_{3}$/Py, and (c) Py/Cu/$\gamma-$IrMn$_{3}$, respectively. A positive exchange bias of 158 Oe was observed for the sample with $\gamma-$IrMn$_{3}$ grown directly on top of Py. STFMR measurement for (d) Py/$\gamma-$IrMn$_{3}$, (e) $\gamma-$IrMn$_{3}$/Py, and (f) Py/Cu/$\gamma-$IrMn$_{3}$, respectively. The solid line is the Lorentzian fit using Eq~\ref{Vdc}. The symmetric ($V_{\rm S}$) and anti-symmetric ($V_{\rm A}$) components of the fitted curve are depicted in black and red, respectively. The STFMR measurements reveal that Py/$\gamma-$IrMn$_{3}$ and $\gamma-$IrMn$_{3}$/Py systems show a dramatic change of lineshape with field reversal, which is absent in the Py/Cu/$\gamma-$IrMn$_{3}$ system.}
\label{fig:2}
\end{figure*}

Figure~\ref{fig:2} shows the magneto-optic Kerr effect (MOKE) measurements performed in the longitudinal configuration with an intensity stabilized He-Ne laser (5 mW and wavelength of 632.8~nm). The incident laser light was \textit{p-}polarized, and the reﬂected light is passed through a photoelastic modulator (PEM) followed by an analyzer kept at 45$^{\circ}$ with respect to the
ﬁrst polarizer. A photodiode was then used to measure the MOKE signal using the lock-in technique. The lock-in ampliﬁer was locked to the second harmonic of PEM frequency. More details of the MOKE setup can be found elsewhere.~\cite{bansal2017crystalline} Figure~\ref{fig:2}(a), (b), and (c) shows the hysteresis loops measured using the magneto-optic Kerr effect at room temperature for Py/$\gamma-$IrMn$_{3}$, $\gamma-$IrMn$_{3}$/Py and Py/Cu/$\gamma-$IrMn$_{3}$ samples, respectively. The sample Py/$\gamma-$IrMn$_{3}$ in which $\gamma-$IrMn$_{3}$ is deposited on top of Py shows large exchange bias, $H_{ex}$ of about 158 Oe. However, we do not observe any exchange bias in samples with reversed stack sequence~\cite{khomenko2008exchange}, \textit{i.e.}, $\gamma-$IrMn$_{3}$/Py as well as in the sample with Cu spacer \textit{i.e.}, Py/Cu/$\gamma-$IrMn$_{3}$.  We use 1.5~nm thick Cu spacer to break any interfacial coupling between Py and $\gamma-$IrMn$_{3}$, as shown in Supplementary Fig.~\ref{Supp:1}. Microstrip devices of dimension ($4~\mu\rm m\times12~\mu\rm m$) were fabricated for the STFMR measurements. For the STFMR measurements, a bias tee is utilized to separate the DC (rectified voltage) and RF port (high-frequency RF current) connected to the microstrip via a Ground-signal-Ground probe.~\cite{kumar2021large} The high-frequency RF current was supplied by R$\&$S signal generator (SMB 100A) at a constant RF power of +3~dBm. The RF signal was amplitude modulated with an 86~Hz signal. The rectified voltage V$_{DC}$ was detected using a lock-in amplifier. The magnetic field was applied in the sample plane at an angle $\varphi$ with respect to the current direction ($x-$direction) along the long axis of the microstrip. The angle, $\varphi$, was varied from $(0-360^{\circ})$ with the help of a vector field magnet. Figure~\ref{fig:2}(c), (d) and (e) show examples of STFMR spectra measured at 8~GHz with $\varphi =~60^{\circ}$ for Py/$\gamma-$IrMn$_{3}$, $\gamma-$IrMn$_{3}$/Py and Py/Cu/$\gamma-$IrMn$_{3}$, respectively. Due to the presence of exchange bias, the resonance field ($H_R$) for the positive and negative fields was found to be different only for Py/$\gamma-$IrMn$_{3}$. We determine the exchange bias as $H_{ex}= |H_R(H>0)-H_R(H<0)|/2$ and found it to be 80~Oe (at $\varphi =~60^{\circ}$), which agrees with the magneto-optic Kerr effect measurements after taking into account of the angular dependence of exchange bias. The lineshape of measured STFMR spectra for the three devices shows dramatically different behavior. For both Py/$\gamma-$IrMn$_{3}$ and $\gamma-$IrMn$_{3}$/Py, the lineshape is found to change upon reversal of the field polarity. However, no such change of lineshape with field polarity is observed for Py/Cu/$\gamma-$IrMn$_{3}$, which is similar to the case of the FM/HM system.~\cite{Liu2011, Macneil2017NP_WTe2} 
The measured STFMR signal ($V_{DC}$) can be fitted using a combination of symmetric and anti-symmetric Lorentzian function~\cite{Liu2011,kumar2021large}:
 
\begin{equation}\label{Vdc}
V_{DC} = V_{S} \frac{\Delta H^2}{\Delta H^2+\small(H-H_{\rm R})^2}\\+V_{A} \frac{\Delta H\small(H-H_{\rm R})}{\Delta H^2+\small(H-H_{\rm R})^2},
\end{equation}

Where $V_{S}$ and $V_{A}$ are symmetric and anti-symmetric voltage components of the rectified voltage. $\triangle H$ represents the linewidth.
The measured STFMR data are fitted with the above equation, as shown by solid lines. We also separately show the symmetric (black) and antisymmetric (red) components. 

The magnitude and/or sign of $V_{S}$ and $V_{A}$ are found to change with the reversal of field polarity for Py/$\gamma-$IrMn$_{3}$ and $\gamma-$IrMn$_{3}$/Py. For a conventional SOT, only sign of $V_{S}$ and $V_{A}$ should change with the reversal of field polarity. Hence, the SOT for Py/$\gamma-$IrMn$_{3}$ and $\gamma-$IrMn$_{3}$/Py can be considered unconventional. However, when the 1.5~nm thick Cu spacer is present between Py and $\gamma-$IrMn$_{3}$, only sign of $V_{S}$ and $V_{A}$ changes with the reversal of field polarity- indicating the absence of USOT. Thus, these results indicate that the USOT in our system does not arise from the bulk $\gamma-$IrMn$_{3}$ but from the interface of Py and $\gamma-$IrMn$_{3}$. This is in stark contrast to the recent results of USOT in L1$_2-$ IrMn$_{3}$/Py by Zhou \textit{et al.}~\cite{zhou2020magnetic}, where the USOT was present even with the insertion of Cu layer. Moreover, Zhou \textit{et al.} used epitaxial L1$_2-$ IrMn$_{3}$, whereas we have $\gamma-$IrMn$_3$ phase, for which no USOT is expected from the broken magnetic mirror symmetry.~\cite{zhou2020magnetic} Additionally, any USOT due to bulk crystal structure is expected to be averaged out for polycrystalline films. Furthermore, USOTs is present even without the exchange bias for $\gamma-$IrMn$_{3}$/Py indicating that USOT, which we observe, is not linked to the exchange bias. 

\begin{figure*} [t!]
\centering
\includegraphics[width=\linewidth]{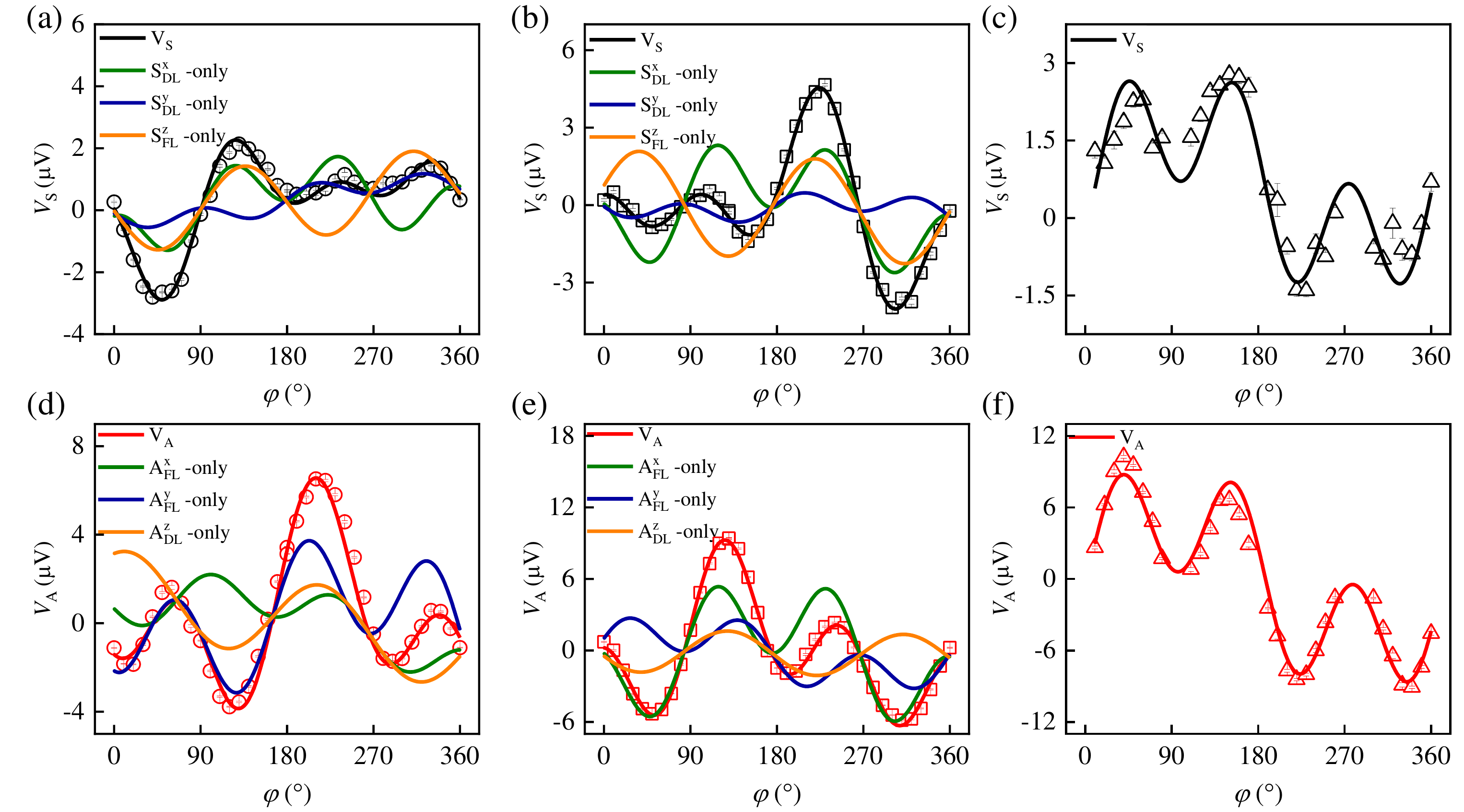}
\caption{Angular dependence of $V_{S}$ and $V_{A}$ components for (a), and (d) Py/$\gamma-$IrMn$_{3}$, (b), and (e) $\gamma-$IrMn$_{3}$/Py (c), and (f) Py/Cu/$\gamma-$IrMn$_{3}$ sample with Cu spacer to avoid exchange coupling between Py and $\gamma-$IrMn$_{3}$. The fitting (solid lines) is performed using Eq.~\ref{VS} and Eq.~\ref{VA}. The other solid lines (brown, blue, and green) show the extracted angular dependence of voltage contributions from $x-$, $y-$, and $z-$direction spin polarization.}
\label{fig:3} 
\end{figure*} 

To systematically examine the presence of USOT and quantify the damping-like torque efficiencies, we have performed complete angular dependent STFMR measurements as shown in Fig.~\ref{fig:3}. 
First, second and third column corresponds to  Py/$\gamma-$IrMn$_{3}$, $\gamma-$IrMn$_{3}$/Py and  Py/Cu/$\gamma-$IrMn$_{3}$, respectively. Figure~\ref{fig:3}(a), (b), and (c) show the extracted symmetric voltage components, and Fig.~\ref{fig:3}(d), (e), and (f) show the extracted antisymmetric voltage components. Conventional SOTs (spin polarization along the \textit{y-}direction) lead to $\rm sin(2\varphi) cos(\varphi)$-dependence for both symmetric and anti-symmetric voltage components.~\cite{Macneil2017NP_WTe2,bose2022tilted} However, the data for Py/$\gamma-$IrMn$_{3}$ and $\gamma-$IrMn$_{3}$/Py do not follow the $\rm sin(2\varphi) cos(\varphi)$-dependence indicating that the spin polarization is not restricted to the $y-$ direction only but also have $x-$ and $z-$ components.
Considering the spin polarization in an arbitrary direction, the angular dependence of  $V_S$ and $V_A$ components can be generalized as~\cite{Macneil2017NP_WTe2,nan2020controlling,bose2022tilted}:

\begin{equation}\label{VS}
V_S = S_{DL}^{x}{\rm sin(2\varphi) sin(\varphi)} + S_{DL}^{y}{\rm sin(2\varphi) cos(\varphi)}+ S_{FL}^{z}\rm sin(2\varphi)
\end{equation}
\begin{equation}\label{VA}
V_A = A_{FL}^{x}{\rm sin(2\varphi) \rm sin(\varphi)} + A_{FL}^{y}{\rm sin(2\varphi) cos(\varphi)}+ A_{DL}^{z}{\rm sin(2\varphi)}.
\end{equation}

Here, $S_{DL}^{x}$, $S_{DL}^{y}$ and $A_{DL}^{z}$ are the coefficients for the damping-like torque generated by the different components of spin Hall conductivity tensor, $\sigma_{zx}^{x}$, $\sigma_{zx}^{y}$, and $\sigma_{zx}^{z}$ respectively. 
Similarly, $A_{FL}^{x}$, $A_{FL}^{y}$, and $S_{FL}^{z}$ represent the coefficients corresponding to a field-like component of the spin Hall conductivity tensor.
Thus by fitting the angular dependence of $V_{S}$ and $V_{A}$ using Eq.~\ref{VS} and Eq.~\ref{VA}, various contributions can be separated. Assuming that $A_{FL}^{y}$ is due to the Oersted field alone, we can quantify the amplitude of three components of damping-like torque efficiencies.~\cite{zhou2020magnetic,bose2022tilted}
\begin{equation}\label{eqSHA_x}
\xi_{\rm DL}^{x} = \frac{S_{DL}^{x}}{A_{FL}^{y}}\frac{e\mu_{0}M_{\rm S}t_{\rm FM}d_{\rm NM}}{\hbar}\left[1+\frac{4\pi M_{\rm eff}}{H_{\rm R}}\right]^{1/2}
\end{equation}

\begin{equation}\label{eqSHA_y}
\xi_{\rm DL}^{y} = \frac{S_{DL}^{y}}{A_{FL}^{y}}\frac{e\mu_{0}M_{\rm S}t_{\rm FM}d_{\rm NM}}{\hbar}\left[1+\frac{4\pi M_{\rm eff}}{H_{\rm R}}\right]^{1/2}
\end{equation}

\begin{equation}\label{eqSHA_z}
\xi_{\rm DL}^{z} = \frac{A_{DL}^{z}}{A_{FL}^{y}}\frac{e\mu_{0}M_{\rm S}t_{\rm FM}d_{\rm NM}}{\hbar}
\end{equation}

Here, $d_{\rm NM}$ and $t_{\rm FM}$ are the non-magnetic and ferromagnetic layer thicknesses, respectively. The parameters $e$, $\hbar$, and $M_{S}$ represent the electric charge, reduced Planck's constant, and saturation magnetization, respectively. The effective magnetization, $M_{\rm eff}$, is determined from the fitting of $f$ versus $H_{R}$ [obtained from Eq.~\ref{Vdc}] data using the Kittel equation.~\cite{Kittle1948}


\begin{table*}[t!]
\caption{Exchange bias, damping-like efficiency, and spin polarization for Py/$\gamma-$IrMn$_{3}$, $\gamma-$IrMn$_{3}$/Py and Py/Cu/$\gamma-$IrMn$_{3}$.}
\centering 

\begin{tabular}{p{3 cm} p{2.5 cm} p{4cm} p{2.5cm} p{2.5cm} p{2.5cm}} 

\hline
Sample & Exchange Bias & Direction of polarization & \multicolumn{3}{c}{Damping-like efficiency} \\
 & (Oe) & & $\xi_{\rm DL}^{x}$ & $\xi_{\rm DL}^{y}$ & $\xi_{\rm DL}^{z}$\\

\hline 
Py/$\gamma-$IrMn$_{3}$ & 158 Oe & $x-$, $y-$, and $z-$ direction & 0.28$\pm$0.04 & 0.08$\pm$0.03 & -0.12$\pm$0.02 \\ 
$\gamma-$IrMn$_{3}$/Py & 0 Oe &$x-$, $y-$, and $z-$ direction & 0.55$\pm$0.07 & -0.25$\pm$0.03 & -0.40$\pm$0.11 \\ 
Py/Cu/$\gamma-$IrMn$_{3}$ & 0 Oe & $y-$ direction only & 0 & 0.16$\pm$0.02 & 0 \\ 
\hline\hline 
\end{tabular}
\label{tabp}
\end{table*}
Table~\ref{tabp} summarizes the results of fitting of angular dependence of $V_{S}$ and $V_{A}$ for
Py/$\gamma-$IrMn$_{3}$, $\gamma-$IrMn$_{3}$/Py and Py/Cu/$\gamma-$IrMn$_{3}$ structure.  
In the table, $\xi_{\rm DL}^{y}$ denotes the effective conventional damping-like efficiency in the $y-$direction, while $\xi_{\rm DL}^{x}$ and $\xi_{\rm DL}^{z}$ represent the unconventional damping-like efficiencies in the $x-$ and $z-$direction, respectively. All the three components $\xi_{\rm DL}^{y}$, $\xi_{\rm DL}^{x}$, and $\xi_{\rm DL}^{z}$ are found to be present in Py/$\gamma-$IrMn$_{3}$, $\gamma-$IrMn$_{3}$/Py system, while only $\xi_{\rm DL}^{y}$ is found to be present in Py/Cu/$\gamma-$IrMn$_{3}$. The conventional damping-like efficiency $\xi_{\rm DL}^{y}$ is found to be (0.08 $\pm$ 0.03), (-0.25 $\pm$ 0.02) and (0.16 $\pm$ 0.02) for Py/$\gamma-$IrMn$_{3}$, $\gamma-$IrMn$_{3}$/Py and Py/Cu/$\gamma-$IrMn$_{3}$, respectively. Here, the sign change in amplitude for $\xi_{\rm DL}^{y}$ for Py/$\gamma-$IrMn$_{3}$, $\gamma-$IrMn$_{3}$/Py is observed because of the stack sequence.
The uncertainty in damping-like efficiency values denotes device-to-device variability and fitting errors in $V_{S}$ and $V_{A}$. The magnitude of $\xi_{\rm DL}^{y}$ for Py/Cu/$\gamma-$IrMn$_{3}$  is found to be greater than that of Py/$\gamma-$IrMn$_{3}$ and lower than that of $\gamma-$IrMn$_{3}$/Py  system.  This implies that the unconventional contributions in Py/$\gamma-$IrMn$_{3}$ and $\gamma-$IrMn$_{3}$/Py directly increase or decrease the bulk $\xi_{\rm DL}^{y}$ resulting in higher or lower effective $\xi_{\rm DL}^{y}$. In an earlier report by Tshitoyan, \textit{et al.}~\cite{tshitoyan2015electrical}, $\xi_{\rm DL}^{y}$ was reported to be higher for Py/IrMn compared to Py/Cu/IrMn, which was correlated to the exchange bias. In contrast, our study shows a higher value of the $\xi_{\rm DL}^{y}$ for Py/Cu/$\gamma-$IrMn$_{3}$ compared to the exchange biased sample Py/$\gamma-$IrMn$_{3}$ indicating a different mechanism of damping-like spin-orbit torque in our system.
We believe this is due to a combination of bulk and interfacial effects.~\cite{amin2020interfacial} The bulk effect from $\gamma-$IrMn$_{3}$ can be considered as the value of $\xi_{\rm DL}^{y}$ we measure for Py/Cu/$\gamma-$IrMn$_{3}$. The interfacial contributions for Py/$\gamma-$IrMn$_{3}$ and $\gamma-$IrMn$_{3}$/Py have opposite signs, which explains the observed values of $\xi_{\rm DL}^{y}$. 

The unconventional efficiencies in our system are of pure interfacial origin, as stated earlier. The $x-$ component of USOT ($\xi_{\rm DL}^{x}$) is found to be 0.28 $\pm$ 0.04 and 0.55 $\pm$ 0.07 for Py/$\gamma-$IrMn$_{3}$ and $\gamma-$IrMn$_{3}$/Py, respectively. Similarly, the $z-$ component of USOT ($\xi_{\rm DL}^{z}$) is found to be -0.12 $\pm$ 0.02 and -0.40 $\pm$ 0.11 for Py/$\gamma-$IrMn$_{3}$ and $\gamma-$IrMn$_{3}$/Py, respectively. Interestingly, unconventional torque efficiencies are larger than conventional $\xi_{\rm DL}^{y}$, which has never been reported before to our knowledge. We also investigated the dependency of USOT on the thickness of Py layer. We do not observe any systematic variation of USOT with Py thickness, as shown in Supplementary Fig.~\ref{Supp:2}. However, we consistently find that the unconventional torque efficiencies are larger than the conventional $\xi_{\rm DL}^{y}$ for a different devices and with varying Py thicknesses.

We believe the generation of USOTs at the interface of Py and $\gamma-$IrMn$_{3}$ could be due to the spin swapping effect~\cite{lifshits2009swapping}. The Py/$\gamma-$IrMn$_{3}$ interface works as a magnetic scattering center causing spin swapping (exchange of spin polarization and flow) of bulk spin currents generated by the interior of $\gamma-$IrMn$_{3}$. The interfacial scattering can deflect the spin current flow, resulting in the generation of spin currents polarized in $x-$ and $z-$ directions, which are non-trivial in nature for normal heavy metals exhibiting the bulk spin Hall effect.~\cite{Manchon2019review} A similar effect was recently observed in another collinear antiferromagnet, Mn$_{3}$Sn.~\cite{hazra2022generation} Hence, we believe the interfacial USOT that we observe is unique to $\gamma-$IrMn$_{3}$, since such USOT has not been reported in IrMn previously. ~\cite{tshitoyan2015electrical, saglam2018independence, zhou2019large}


We observed unconventional spin-orbit torques in heterostructures consisting of Py and polycrystalline $\gamma-$IrMn$_{3}$. The systematic investigation of the angular dependence of symmetric and anti-symmetric components of STFMR measurements suggests the presence of unconventional spin-orbit torques in Py/$\gamma-$IrMn$_{3}$ and $\gamma-$IrMn$_{3}$/Py. The interface between Py and $\gamma-$IrMn$_{3}$ gives rise to substantial out-of-plane damping-like spin-orbit torques in addition to the conventional spin-orbit torques. These unconventional spin-orbit torques can be eliminated by inserting a Cu spacer between Py and $\gamma-$IrMn$_{3}$-indicating an interfacial origin of the unconventional spin-orbit torques in contrast to the previously reported studies that demonstrated a bulk origin arising from the magnetic mirror asymmetry in IrMn$_{3}$. A large in-plane damping-like efficiency of 0.16 is also observed for the Py/Cu/$\gamma-$IrMn$_{3}$ system, indicating that $\gamma-$IrMn$_{3}$ is an efficient source of spin current. These findings emphasize the importance of investigating the interfacial origin of these torques to comprehend the physics of collinear antiferromagnets. The large out-of-plane damping-like efficiency that we observed in our system can be utilized to effectively switch a perpendicularly magnetized system used in high-density magnetic recording.~\cite{Liu2012, fukami2016magnetization} Therefore, it has high technological relevance for energy-efficient spintronic devices based on quantum materials.

\begin{acknowledgments}
We acknowledge the Department of Science and Technology for supporting this collaborative work under the Indo-Japan project (grant no: $DST/INT/JSPS/P-266/2018$). The partial support from the Science \& Engineering Research Board (SERB File no. CRG/2018/001012), the Ministry of Human Resource Development under the IMPRINT program (Grant no: 7519 and 7058), the Department of Science and Technology under the Nanomission program (grant no: $SR/NM/NT-1041/2016(G)$), the Department of Electronics and Information Technology (DeitY), Joint Advanced Technology Centre at IIT Delhi, and the Grand Challenge project supported by IIT Delhi are gratefully acknowledged. 
A.~K. acknowledges support from the Council of Scientific and Industrial Research (CSIR), India.
\end{acknowledgments}

\section{AUTHOR DECLARATIONS}
\subsection*{Conflict of Interest}
The authors have no conflicts to disclose.

\section{DATA AVAILABILITY}
The data that support the findings of this study are available from the corresponding author upon reasonable request.
\providecommand{\noopsort}[1]{}\providecommand{\singleletter}[1]{#1}%

\begin{center}
\textbf{\large Supplemental Materials: Interfacial origin of unconventional spin-orbit torque in Py/$\gamma-$IrMn$_{3}$}
\end{center}
\setcounter{equation}{0}
\setcounter{figure}{0}
\setcounter{table}{0}
\setcounter{page}{1}
\makeatletter
\renewcommand{\theequation}{S\arabic{equation}}
\renewcommand{\thefigure}{S\arabic{figure}}
\renewcommand{\bibnumfmt}[1]{[S#1]}
\renewcommand{\citenumfont}[1]{S#1}
\renewcommand{\thetable}{S\arabic{table}}
\setcounter{section}{0}
\renewcommand{\thesection}{S-\Roman{section}}

\section{Effect of Cu thickness on interface between Py and $\gamma-$IrMn$_{3}$}

\begin{figure*} [btp!]
\centering
\includegraphics[width=0.7\linewidth]{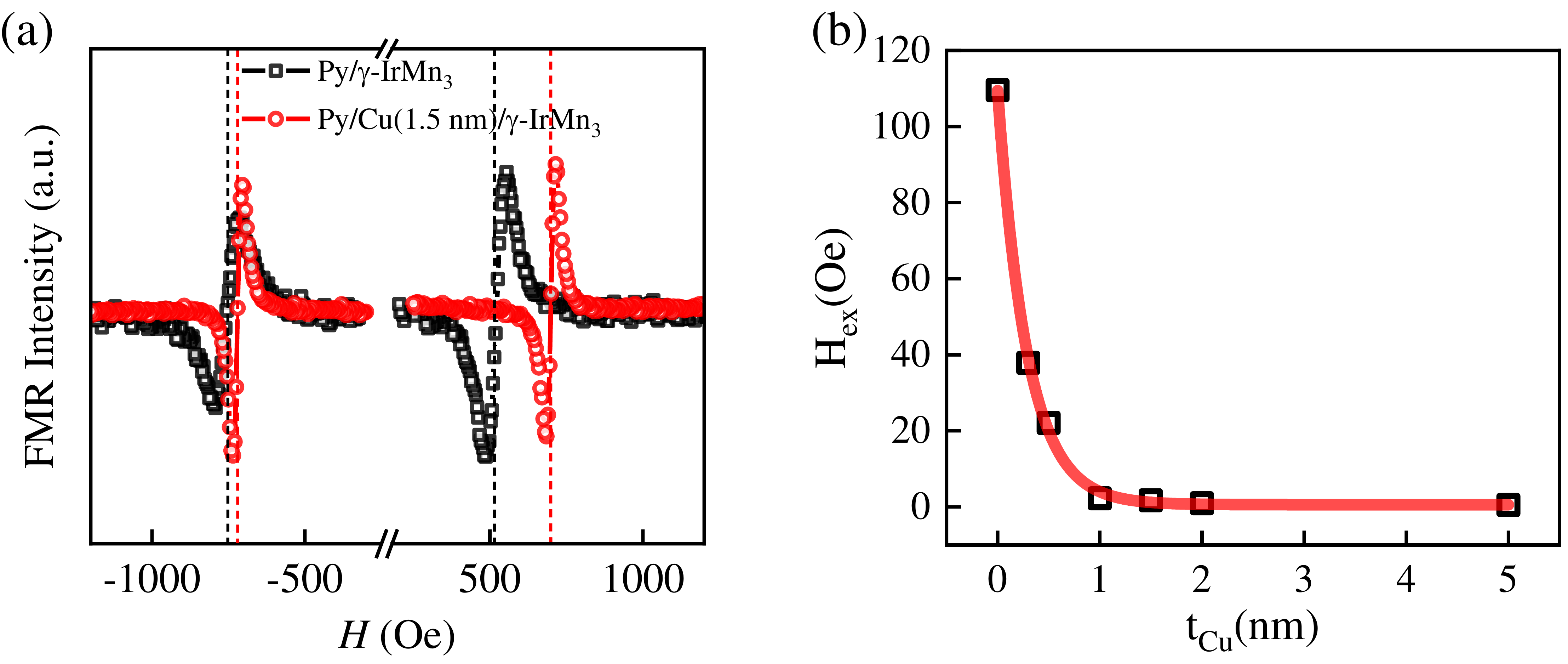}
\caption{(a)FMR measurements for Py/$\gamma-$IrMn$_{3}$ (black) and Py/Cu (1.5~nm)/$\gamma-$IrMn$_{3}$ (red) sample measured at 8~GHz with sweeping magnetic field parallel to exchange bias. The field reversal spectra show that exchange bias generates a shift in the resonance field in Py/$\gamma-$IrMn$_{3}$. (b) Variation of exchange bias as a function of Cu spacer thickness.}
\label{Supp:1} 
\end{figure*}

 A Cu spacer can control the exchange coupling in the Py/$\gamma-$IrMn$_{3}$ thin films. We use the ferromagnetic resonance (FMR) method to determine the effect of Cu thickness at the Py/$\gamma-$IrMn$_{3}$ interface. From FMR measurements, we determine the exchange bias of  Py/$\gamma-$IrMn$_{3}$ from the resonance fields ($H_R$) for the positive and negative fields. 
 Figure~\ref{Supp:1}(a) shows measured FMR spectra for Py/$\gamma-$IrMn$_{3}$ and Py/Cu/$\gamma-$IrMn$_{3}$  samples. The resonance field in Py/$\gamma-$IrMn$_{3}$ is changed by reversing the field direction.
 This is caused by the exchange bias, which can be determined as $H_{ex}= |H_R(H>0)-H_R(H<0)|/2$. Figure~\ref{Supp:1}(b) shows the behavior of H$_{ex}$ with the thickness of the Cu spacer. The exchange bias shows a steep decrease with the thickness of the Cu layer. The exchange bias completely disappears after a thickness of 1~nm of Cu. Hence, we chose Py/Cu/IrMn$_{3}$ samples with 1.5~nm Cu in the main text to eliminate the exchange coupling between Py and $\gamma-$IrMn$_{3}$. 

\section{Thickness dependence of the SOT efficiencies}

\begin{figure*} [btp!]
\centering
\includegraphics[width=\linewidth]{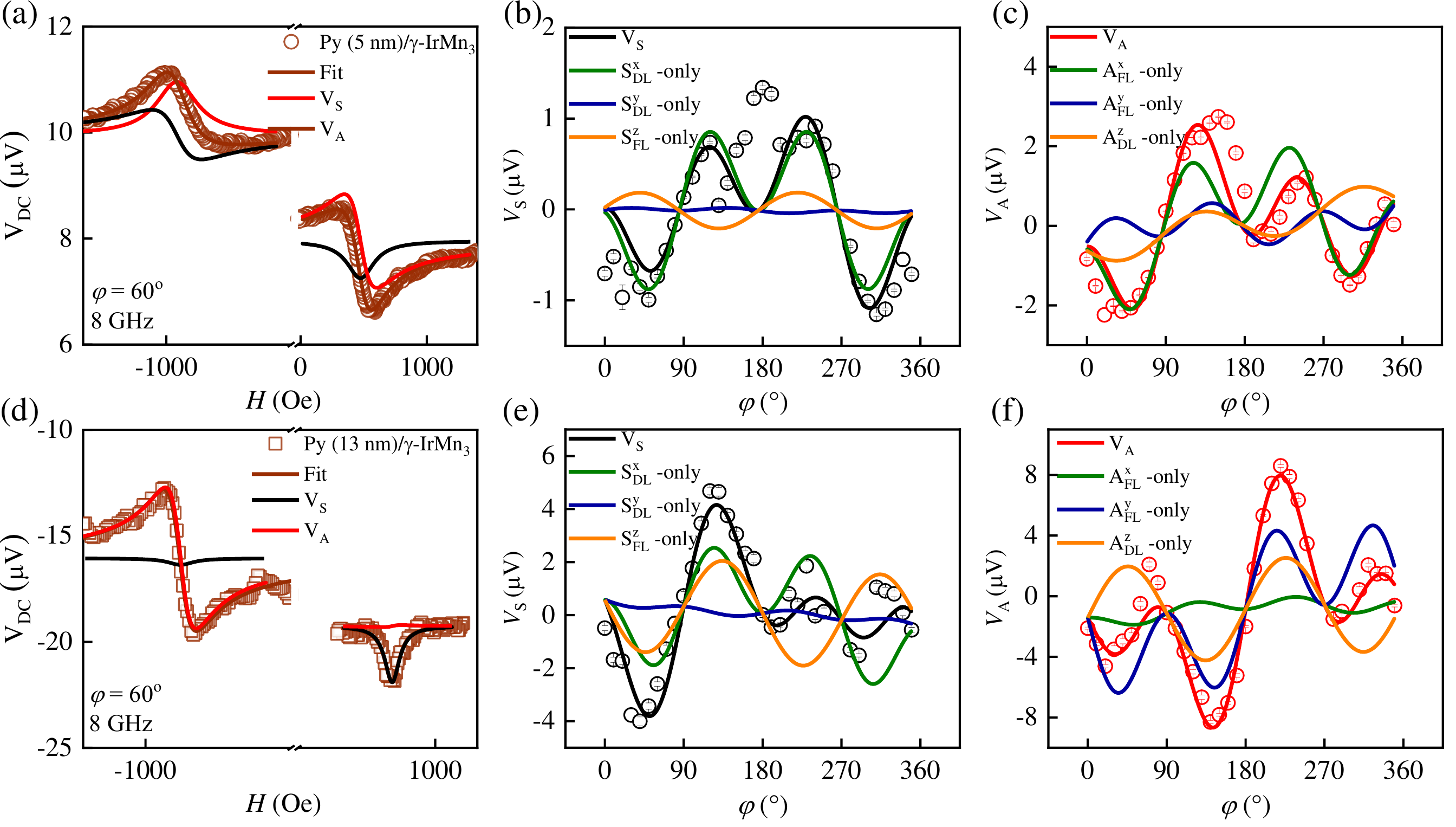}
\caption{STFMR measurements on (a) Py(5 nm)/$\gamma-$IrMn$_{3}$, and (d) Py(13 nm)/$\gamma-$IrMn$_{3}$, Angular dependence of $V_{S}$ and $V_{A}$ components for (b), and (c) Py(5 nm)/$\gamma-$IrMn$_{3}$, and (e), and (f) Py(13 nm)/$\gamma-$IrMn$_{3}$. The solid lines in (a) and (d) represent Lorentzian fit using Eq (1) of the main text. The symmetric ($V_{\rm S}$) and anti-symmetric ($V_{\rm A}$) components of the fitted curve are depicted in black and red, respectively. The fitting (solid lines) in (b), (c), (e), and (f) are performed using Eq.~(2) and Eq.~(3) of the main text. The  solid lines (brown, blue, and green) in these figures show the extracted angular dependence of voltage contributions from $x-$, $y-$, and $z-$direction spin polarization.}
\label{Supp:2} 
\end{figure*}

SOT efficiency was measured for different thicknesses of Py in contact with $\gamma-$IrMn$_{3}$. Figure~\ref{Supp:2} (a), (d) shows the STFMR spectra for Py(5~nm)/$\gamma-$IrMn$_{3}$ and Py(13~nm)/$\gamma-$IrMn$_{3}$. The angular-dependent of V$_S$ and V$_A$ is shown in Fig.~\ref{Supp:2} (b), (c), (e), and (f) show the presence of unconventional torque in both systems. We summarize the extracted torques in Table~S1. Though all samples with Py/$\gamma-$IrMn$_{3}$ show the presence of USOT, no systematic thickness dependence is observed by varying Py thickness from 5~nm$\rightarrow$8.7~nm$\rightarrow$13~nm. 
We also found larger $\xi_{\rm DL}^{z}$ $>$ $\xi_{\rm DL}^{y}$ for all the samples.

\begin{table*}[h!]
\caption{Spin polarization direction and damping-like efficiency for different Py thickness: }
\centering 

\begin{tabular}{p{4 cm}  p{5cm} p{2cm} p{2cm} p{2cm}} 

\hline
Sample  & Direction of polarization & \multicolumn{3}{c}{Damping-like efficiency} \\
 &  & $\xi_{\rm DL}^{x}$ & $\xi_{\rm DL}^{y}$ & $\xi_{\rm DL}^{z}$\\

\hline 
Py (5 nm)/$\gamma-$IrMn$_{3}$  & $x-$, $y-$, and $z-$ direction & -0.58$\pm$0.19 & 0.02$\pm$0.10 & -0.06$\pm$0.04 \\ 
Py (13 nm)/$\gamma-$IrMn$_{3}$  &$x-$, $y-$, and $z-$ direction & 0.53$\pm$0.06 & 0.04$\pm$0.08 & -0.15$\pm$0.02 \\ 
\hline\hline 
\end{tabular}
\label{tabp}
\end{table*}


\begin{thebibliography}{52}%
\makeatletter
\providecommand \@ifxundefined [1]{%
 \@ifx{#1\undefined}
}%
\providecommand \@ifnum [1]{%
 \ifnum #1\expandafter \@firstoftwo
 \else \expandafter \@secondoftwo
 \fi
}%
\providecommand \@ifx [1]{%
 \ifx #1\expandafter \@firstoftwo
 \else \expandafter \@secondoftwo
 \fi
}%
\providecommand \natexlab [1]{#1}%
\providecommand \enquote  [1]{``#1''}%
\providecommand \bibnamefont  [1]{#1}%
\providecommand \bibfnamefont [1]{#1}%
\providecommand \citenamefont [1]{#1}%
\providecommand \href@noop [0]{\@secondoftwo}%
\providecommand \href [0]{\begingroup \@sanitize@url \@href}%
\providecommand \@href[1]{\@@startlink{#1}\@@href}%
\providecommand \@@href[1]{\endgroup#1\@@endlink}%
\providecommand \@sanitize@url [0]{\catcode `\\12\catcode `\$12\catcode
  `\&12\catcode `\#12\catcode `\^12\catcode `\_12\catcode `\%12\relax}%
\providecommand \@@startlink[1]{}%
\providecommand \@@endlink[0]{}%
\providecommand \url  [0]{\begingroup\@sanitize@url \@url }%
\providecommand \@url [1]{\endgroup\@href {#1}{\urlprefix }}%
\providecommand \urlprefix  [0]{URL }%
\providecommand \Eprint [0]{\href }%
\providecommand \doibase [0]{http://dx.doi.org/}%
\providecommand \selectlanguage [0]{\@gobble}%
\providecommand \bibinfo  [0]{\@secondoftwo}%
\providecommand \bibfield  [0]{\@secondoftwo}%
\providecommand \translation [1]{[#1]}%
\providecommand \BibitemOpen [0]{}%
\providecommand \bibitemStop [0]{}%
\providecommand \bibitemNoStop [0]{.\EOS\space}%
\providecommand \EOS [0]{\spacefactor3000\relax}%
\providecommand \BibitemShut  [1]{\csname bibitem#1\endcsname}%
\let\auto@bib@innerbib\@empty
\bibitem [{\citenamefont {Ando}\ \emph {et~al.}(2008)\citenamefont {Ando},
  \citenamefont {Takahashi}, \citenamefont {Harii}, \citenamefont {Sasage},
  \citenamefont {Ieda}, \citenamefont {Maekawa},\ and\ \citenamefont
  {Saitoh}}]{Ando2008}%
  \BibitemOpen
  \bibfield  {author} {\bibinfo {author} {\bibfnamefont {K.}~\bibnamefont
  {Ando}}, \bibinfo {author} {\bibfnamefont {S.}~\bibnamefont {Takahashi}},
  \bibinfo {author} {\bibfnamefont {K.}~\bibnamefont {Harii}}, \bibinfo
  {author} {\bibfnamefont {K.}~\bibnamefont {Sasage}}, \bibinfo {author}
  {\bibfnamefont {J.}~\bibnamefont {Ieda}}, \bibinfo {author} {\bibfnamefont
  {S.}~\bibnamefont {Maekawa}}, \ and\ \bibinfo {author} {\bibfnamefont
  {E.}~\bibnamefont {Saitoh}},\ }\href {\doibase
  10.1103/PhysRevLett.101.036601} {\bibfield  {journal} {\bibinfo  {journal}
  {Phys.\ Rev.\ Lett.}\ }\textbf {\bibinfo {volume} {101}},\ \bibinfo {pages}
  {036601} (\bibinfo {year} {2008})}\BibitemShut {NoStop}%
\bibitem [{\citenamefont {Gambardella}\ and\ \citenamefont
  {Miron}(2011)}]{gambardella2011current}%
  \BibitemOpen
  \bibfield  {author} {\bibinfo {author} {\bibfnamefont {P.}~\bibnamefont
  {Gambardella}}\ and\ \bibinfo {author} {\bibfnamefont {I.~M.}\ \bibnamefont
  {Miron}},\ }\href
  {https://royalsocietypublishing.org/doi/10.1098/rsta.2010.0336} {\bibfield
  {journal} {\bibinfo  {journal} {Proc. Natl. Acad. Sci.}\ }\textbf {\bibinfo
  {volume} {369}},\ \bibinfo {pages} {3175} (\bibinfo {year}
  {2011})}\BibitemShut {NoStop}%
\bibitem [{\citenamefont {{Liu}}\ \emph {et~al.}(2012)\citenamefont {{Liu}},
  \citenamefont {{Pai}}, \citenamefont {{Li}}, \citenamefont {{Tseng}},
  \citenamefont {{Ralph}},\ and\ \citenamefont {{Buhrman}}}]{Liu2012}%
  \BibitemOpen
  \bibfield  {author} {\bibinfo {author} {\bibfnamefont {L.}~\bibnamefont
  {{Liu}}}, \bibinfo {author} {\bibfnamefont {C.-F.}\ \bibnamefont {{Pai}}},
  \bibinfo {author} {\bibfnamefont {Y.}~\bibnamefont {{Li}}}, \bibinfo {author}
  {\bibfnamefont {H.~W.}\ \bibnamefont {{Tseng}}}, \bibinfo {author}
  {\bibfnamefont {D.~C.}\ \bibnamefont {{Ralph}}}, \ and\ \bibinfo {author}
  {\bibfnamefont {R.~A.}\ \bibnamefont {{Buhrman}}},\ }\href {\doibase
  10.1126/science.1218197} {\bibfield  {journal} {\bibinfo  {journal}
  {Science}\ }\textbf {\bibinfo {volume} {336}},\ \bibinfo {pages} {555}
  (\bibinfo {year} {2012})}\BibitemShut {NoStop}%
\bibitem [{\citenamefont {Demidov}\ \emph {et~al.}(2012)\citenamefont
  {Demidov}, \citenamefont {Urazhdin}, \citenamefont {Ulrichs}, \citenamefont
  {Tiberkevich}, \citenamefont {Slavin}, \citenamefont {Baither}, \citenamefont
  {Schmitz},\ and\ \citenamefont {Demokritov}}]{VEDemidov2012}%
  \BibitemOpen
  \bibfield  {author} {\bibinfo {author} {\bibfnamefont {V.~E.}\ \bibnamefont
  {Demidov}}, \bibinfo {author} {\bibfnamefont {S.}~\bibnamefont {Urazhdin}},
  \bibinfo {author} {\bibfnamefont {H.}~\bibnamefont {Ulrichs}}, \bibinfo
  {author} {\bibfnamefont {V.}~\bibnamefont {Tiberkevich}}, \bibinfo {author}
  {\bibfnamefont {A.}~\bibnamefont {Slavin}}, \bibinfo {author} {\bibfnamefont
  {D.}~\bibnamefont {Baither}}, \bibinfo {author} {\bibfnamefont
  {G.}~\bibnamefont {Schmitz}}, \ and\ \bibinfo {author} {\bibfnamefont
  {S.~O.}\ \bibnamefont {Demokritov}},\ }\href {\doibase 10.1038/nmat3459}
  {\bibfield  {journal} {\bibinfo  {journal} {Nat.\ Mater.}\ }\textbf {\bibinfo
  {volume} {11}},\ \bibinfo {pages} {1028} (\bibinfo {year}
  {2012})}\BibitemShut {NoStop}%
\bibitem [{\citenamefont {{Emori}}\ \emph {et~al.}(2013)\citenamefont
  {{Emori}}, \citenamefont {{Bauer}}, \citenamefont {{Ahn}}, \citenamefont
  {{Martinez}},\ and\ \citenamefont {{Beach}}}]{Emori2013}%
  \BibitemOpen
  \bibfield  {author} {\bibinfo {author} {\bibfnamefont {S.}~\bibnamefont
  {{Emori}}}, \bibinfo {author} {\bibfnamefont {U.}~\bibnamefont {{Bauer}}},
  \bibinfo {author} {\bibfnamefont {S.-M.}\ \bibnamefont {{Ahn}}}, \bibinfo
  {author} {\bibfnamefont {E.}~\bibnamefont {{Martinez}}}, \ and\ \bibinfo
  {author} {\bibfnamefont {G.~S.~D.}\ \bibnamefont {{Beach}}},\ }\href
  {\doibase 10.1038/nmat3675} {\bibfield  {journal} {\bibinfo  {journal} {Nat.\
  Mater.}\ }\textbf {\bibinfo {volume} {12}},\ \bibinfo {pages} {611} (\bibinfo
  {year} {2013})}\BibitemShut {NoStop}%
\bibitem [{\citenamefont {Awad}\ \emph {et~al.}(2016)\citenamefont {Awad},
  \citenamefont {D{\"{u}}rrenfeld}, \citenamefont {Houshang}, \citenamefont
  {Dvornik}, \citenamefont {Iacocca}, \citenamefont {Dumas},\ and\
  \citenamefont {{\AA}kerman}}]{awad2016natphys}%
  \BibitemOpen
  \bibfield  {author} {\bibinfo {author} {\bibfnamefont {A.~A.}\ \bibnamefont
  {Awad}}, \bibinfo {author} {\bibfnamefont {P.}~\bibnamefont
  {D{\"{u}}rrenfeld}}, \bibinfo {author} {\bibfnamefont {A.}~\bibnamefont
  {Houshang}}, \bibinfo {author} {\bibfnamefont {M.}~\bibnamefont {Dvornik}},
  \bibinfo {author} {\bibfnamefont {E.}~\bibnamefont {Iacocca}}, \bibinfo
  {author} {\bibfnamefont {R.~K.}\ \bibnamefont {Dumas}}, \ and\ \bibinfo
  {author} {\bibfnamefont {J.}~\bibnamefont {{\AA}kerman}},\ }\href {\doibase
  10.1038/nphys3927} {\bibfield  {journal} {\bibinfo  {journal} {Nat. Phys.}\
  }\textbf {\bibinfo {volume} {13}},\ \bibinfo {pages} {292} (\bibinfo {year}
  {2016})}\BibitemShut {NoStop}%
\bibitem [{\citenamefont {Garello}\ \emph {et~al.}(2014)\citenamefont
  {Garello}, \citenamefont {Avci}, \citenamefont {Miron}, \citenamefont
  {Baumgartner}, \citenamefont {Ghosh}, \citenamefont {Auffret}, \citenamefont
  {Boulle}, \citenamefont {Gaudin},\ and\ \citenamefont
  {Gambardella}}]{garello2014ultrafast}%
  \BibitemOpen
  \bibfield  {author} {\bibinfo {author} {\bibfnamefont {K.}~\bibnamefont
  {Garello}}, \bibinfo {author} {\bibfnamefont {C.~O.}\ \bibnamefont {Avci}},
  \bibinfo {author} {\bibfnamefont {I.~M.}\ \bibnamefont {Miron}}, \bibinfo
  {author} {\bibfnamefont {M.}~\bibnamefont {Baumgartner}}, \bibinfo {author}
  {\bibfnamefont {A.}~\bibnamefont {Ghosh}}, \bibinfo {author} {\bibfnamefont
  {S.}~\bibnamefont {Auffret}}, \bibinfo {author} {\bibfnamefont
  {O.}~\bibnamefont {Boulle}}, \bibinfo {author} {\bibfnamefont
  {G.}~\bibnamefont {Gaudin}}, \ and\ \bibinfo {author} {\bibfnamefont
  {P.}~\bibnamefont {Gambardella}},\ }\href
  {https://aip.scitation.org/doi/10.1063/1.4902443} {\bibfield  {journal}
  {\bibinfo  {journal} {Appl.\ Phys.\ Lett.}\ }\textbf {\bibinfo {volume}
  {105}},\ \bibinfo {pages} {212402} (\bibinfo {year} {2014})}\BibitemShut
  {NoStop}%
\bibitem [{\citenamefont {Chen}\ \emph {et~al.}(2016)\citenamefont {Chen},
  \citenamefont {Dumas}, \citenamefont {Eklund}, \citenamefont {Muduli},
  \citenamefont {Houshang}, \citenamefont {Awad}, \citenamefont {Durrenfeld},
  \citenamefont {Malm}, \citenamefont {Rusu},\ and\ \citenamefont
  {{\AA}kerman}}]{chen2016ieeeproc}%
  \BibitemOpen
  \bibfield  {author} {\bibinfo {author} {\bibfnamefont {T.}~\bibnamefont
  {Chen}}, \bibinfo {author} {\bibfnamefont {R.~K.}\ \bibnamefont {Dumas}},
  \bibinfo {author} {\bibfnamefont {A.}~\bibnamefont {Eklund}}, \bibinfo
  {author} {\bibfnamefont {P.~K.}\ \bibnamefont {Muduli}}, \bibinfo {author}
  {\bibfnamefont {A.}~\bibnamefont {Houshang}}, \bibinfo {author}
  {\bibfnamefont {A.~A.}\ \bibnamefont {Awad}}, \bibinfo {author}
  {\bibfnamefont {P.}~\bibnamefont {Durrenfeld}}, \bibinfo {author}
  {\bibfnamefont {B.~G.}\ \bibnamefont {Malm}}, \bibinfo {author}
  {\bibfnamefont {A.}~\bibnamefont {Rusu}}, \ and\ \bibinfo {author}
  {\bibfnamefont {J.}~\bibnamefont {{\AA}kerman}},\ }\href {\doibase
  10.1109/JPROC.2016.2554518} {\bibfield  {journal} {\bibinfo  {journal} {Proc.
  IEEE}\ }\textbf {\bibinfo {volume} {104}},\ \bibinfo {pages} {1919} (\bibinfo
  {year} {2016})}\BibitemShut {NoStop}%
\bibitem [{\citenamefont {Behera}\ \emph {et~al.}(2022)\citenamefont {Behera},
  \citenamefont {Fulara}, \citenamefont {Bainsla}, \citenamefont {Kumar},
  \citenamefont {Zahedinejad}, \citenamefont {Houshang},\ and\ \citenamefont
  {{\AA}kerman}}]{behera2022energy}%
  \BibitemOpen
  \bibfield  {author} {\bibinfo {author} {\bibfnamefont {N.}~\bibnamefont
  {Behera}}, \bibinfo {author} {\bibfnamefont {H.}~\bibnamefont {Fulara}},
  \bibinfo {author} {\bibfnamefont {L.}~\bibnamefont {Bainsla}}, \bibinfo
  {author} {\bibfnamefont {A.}~\bibnamefont {Kumar}}, \bibinfo {author}
  {\bibfnamefont {M.}~\bibnamefont {Zahedinejad}}, \bibinfo {author}
  {\bibfnamefont {A.}~\bibnamefont {Houshang}}, \ and\ \bibinfo {author}
  {\bibfnamefont {J.}~\bibnamefont {{\AA}kerman}},\ }\href {\doibase
  10.1103/PhysRevApplied.18.024017} {\bibfield  {journal} {\bibinfo  {journal}
  {Phys. Rev. Appl.}\ }\textbf {\bibinfo {volume} {18}},\ \bibinfo {pages}
  {024017} (\bibinfo {year} {2022})}\BibitemShut {NoStop}%
\bibitem [{\citenamefont {Kumar}\ \emph {et~al.}(2023)\citenamefont {Kumar},
  \citenamefont {Fulara}, \citenamefont {Khymyn}, \citenamefont {Zahedinejad},
  \citenamefont {Rajabali}, \citenamefont {Zhao}, \citenamefont {Behera},
  \citenamefont {Houshang}, \citenamefont {Awad},\ and\ \citenamefont
  {{\AA}kerman}}]{kumar2023robust}%
  \BibitemOpen
  \bibfield  {author} {\bibinfo {author} {\bibfnamefont {A.}~\bibnamefont
  {Kumar}}, \bibinfo {author} {\bibfnamefont {H.}~\bibnamefont {Fulara}},
  \bibinfo {author} {\bibfnamefont {R.}~\bibnamefont {Khymyn}}, \bibinfo
  {author} {\bibfnamefont {M.}~\bibnamefont {Zahedinejad}}, \bibinfo {author}
  {\bibfnamefont {M.}~\bibnamefont {Rajabali}}, \bibinfo {author}
  {\bibfnamefont {X.}~\bibnamefont {Zhao}}, \bibinfo {author} {\bibfnamefont
  {N.}~\bibnamefont {Behera}}, \bibinfo {author} {\bibfnamefont
  {A.}~\bibnamefont {Houshang}}, \bibinfo {author} {\bibfnamefont {A.~A.}\
  \bibnamefont {Awad}}, \ and\ \bibinfo {author} {\bibfnamefont
  {J.}~\bibnamefont {{\AA}kerman}},\ }\href
  {https://doi.org/10.48550/arXiv.2301.03859} {\bibfield  {journal} {\bibinfo
  {journal} {arXiv}\ }\textbf {\bibinfo {volume} {2301.03859}} (\bibinfo {year}
  {2023})}\BibitemShut {NoStop}%
\bibitem [{\citenamefont {Zahedinejad}\ \emph {et~al.}(2020)\citenamefont
  {Zahedinejad}, \citenamefont {Awad}, \citenamefont {Muralidhar},
  \citenamefont {Khymyn}, \citenamefont {Fulara}, \citenamefont {Mazraati},
  \citenamefont {Dvornik},\ and\ \citenamefont
  {{\AA}kerman}}]{zahedinejad2019two}%
  \BibitemOpen
  \bibfield  {author} {\bibinfo {author} {\bibfnamefont {M.}~\bibnamefont
  {Zahedinejad}}, \bibinfo {author} {\bibfnamefont {A.~A.}\ \bibnamefont
  {Awad}}, \bibinfo {author} {\bibfnamefont {S.}~\bibnamefont {Muralidhar}},
  \bibinfo {author} {\bibfnamefont {R.}~\bibnamefont {Khymyn}}, \bibinfo
  {author} {\bibfnamefont {H.}~\bibnamefont {Fulara}}, \bibinfo {author}
  {\bibfnamefont {H.}~\bibnamefont {Mazraati}}, \bibinfo {author}
  {\bibfnamefont {M.}~\bibnamefont {Dvornik}}, \ and\ \bibinfo {author}
  {\bibfnamefont {J.}~\bibnamefont {{\AA}kerman}},\ }\href
  {https://www.nature.com/articles/s41565-019-0593-9} {\bibfield  {journal}
  {\bibinfo  {journal} {Nat.\ Nano.}\ }\textbf {\bibinfo {volume} {15}},\
  \bibinfo {pages} {47} (\bibinfo {year} {2020})}\BibitemShut {NoStop}%
\bibitem [{\citenamefont {Houshang}\ \emph {et~al.}(2022)\citenamefont
  {Houshang}, \citenamefont {Zahedinejad}, \citenamefont {Muralidhar},
  \citenamefont {Khymyn}, \citenamefont {Rajabali}, \citenamefont {Fulara},
  \citenamefont {Awad}, \citenamefont {{\AA}kerman}, \citenamefont
  {Checi{\'n}ski},\ and\ \citenamefont {Dvornik}}]{houshang2020spin}%
  \BibitemOpen
  \bibfield  {author} {\bibinfo {author} {\bibfnamefont {A.}~\bibnamefont
  {Houshang}}, \bibinfo {author} {\bibfnamefont {M.}~\bibnamefont
  {Zahedinejad}}, \bibinfo {author} {\bibfnamefont {S.}~\bibnamefont
  {Muralidhar}}, \bibinfo {author} {\bibfnamefont {R.}~\bibnamefont {Khymyn}},
  \bibinfo {author} {\bibfnamefont {M.}~\bibnamefont {Rajabali}}, \bibinfo
  {author} {\bibfnamefont {H.}~\bibnamefont {Fulara}}, \bibinfo {author}
  {\bibfnamefont {A.~A.}\ \bibnamefont {Awad}}, \bibinfo {author}
  {\bibfnamefont {J.}~\bibnamefont {{\AA}kerman}}, \bibinfo {author}
  {\bibfnamefont {J.}~\bibnamefont {Checi{\'n}ski}}, \ and\ \bibinfo {author}
  {\bibfnamefont {M.}~\bibnamefont {Dvornik}},\ }\href
  {https://journals.aps.org/prapplied/abstract/10.1103/PhysRevApplied.17.014003}
  {\bibfield  {journal} {\bibinfo  {journal} {Phys.\ Rev.\ Appl.}\ }\textbf
  {\bibinfo {volume} {17}},\ \bibinfo {pages} {014003} (\bibinfo {year}
  {2022})}\BibitemShut {NoStop}%
\bibitem [{\citenamefont {Garg}\ \emph {et~al.}(2021)\citenamefont {Garg},
  \citenamefont {Bhotla}, \citenamefont {Muduli},\ and\ \citenamefont
  {Bhowmik}}]{garg2021kuramoto}%
  \BibitemOpen
  \bibfield  {author} {\bibinfo {author} {\bibfnamefont {N.}~\bibnamefont
  {Garg}}, \bibinfo {author} {\bibfnamefont {S.~V.~H.}\ \bibnamefont {Bhotla}},
  \bibinfo {author} {\bibfnamefont {P.~K.}\ \bibnamefont {Muduli}}, \ and\
  \bibinfo {author} {\bibfnamefont {D.}~\bibnamefont {Bhowmik}},\ }\href
  {https://iopscience.iop.org/article/10.1088/2634-4386/ac3258} {\bibfield
  {journal} {\bibinfo  {journal} {Neuromorph. Comput. Eng.}\ }\textbf {\bibinfo
  {volume} {1}},\ \bibinfo {pages} {024005} (\bibinfo {year}
  {2021})}\BibitemShut {NoStop}%
\bibitem [{\citenamefont {Kumar}\ \emph {et~al.}(2022)\citenamefont {Kumar},
  \citenamefont {Rajabali}, \citenamefont {Gonz{\'a}lez}, \citenamefont
  {Zahedinejad}, \citenamefont {Houshang},\ and\ \citenamefont
  {{\AA}kerman}}]{kumar2022nanoscale}%
  \BibitemOpen
  \bibfield  {author} {\bibinfo {author} {\bibfnamefont {A.}~\bibnamefont
  {Kumar}}, \bibinfo {author} {\bibfnamefont {M.}~\bibnamefont {Rajabali}},
  \bibinfo {author} {\bibfnamefont {V.~H.}\ \bibnamefont {Gonz{\'a}lez}},
  \bibinfo {author} {\bibfnamefont {M.}~\bibnamefont {Zahedinejad}}, \bibinfo
  {author} {\bibfnamefont {A.}~\bibnamefont {Houshang}}, \ and\ \bibinfo
  {author} {\bibfnamefont {J.}~\bibnamefont {{\AA}kerman}},\ }\href {\doibase
  10.1039/D1NR07505E} {\bibfield  {journal} {\bibinfo  {journal} {Nanoscale}\
  }\textbf {\bibinfo {volume} {14}},\ \bibinfo {pages} {1432} (\bibinfo {year}
  {2022})}\BibitemShut {NoStop}%
\bibitem [{\citenamefont {Yadav}\ \emph {et~al.}(2023)\citenamefont {Yadav},
  \citenamefont {Gupta}, \citenamefont {Holla}, \citenamefont {Ali~Khan},
  \citenamefont {Muduli},\ and\ \citenamefont
  {Bhowmik}}]{yadav2023demonstration}%
  \BibitemOpen
  \bibfield  {author} {\bibinfo {author} {\bibfnamefont {R.~S.}\ \bibnamefont
  {Yadav}}, \bibinfo {author} {\bibfnamefont {P.}~\bibnamefont {Gupta}},
  \bibinfo {author} {\bibfnamefont {A.}~\bibnamefont {Holla}}, \bibinfo
  {author} {\bibfnamefont {K.~I.}\ \bibnamefont {Ali~Khan}}, \bibinfo {author}
  {\bibfnamefont {P.~K.}\ \bibnamefont {Muduli}}, \ and\ \bibinfo {author}
  {\bibfnamefont {D.}~\bibnamefont {Bhowmik}},\ }\href
  {https://pubs.acs.org/doi/full/10.1021/acsaelm.2c01488} {\bibfield  {journal}
  {\bibinfo  {journal} {ACS Appl. Electron. Mater.}\ }\textbf {\bibinfo
  {volume} {5}},\ \bibinfo {pages} {484} (\bibinfo {year} {2023})}\BibitemShut
  {NoStop}%
\bibitem [{\citenamefont {{Dyakonov}}\ and\ \citenamefont
  {{Perel}}(1971)}]{Dyakonov1971}%
  \BibitemOpen
  \bibfield  {author} {\bibinfo {author} {\bibfnamefont {M.~I.}\ \bibnamefont
  {{Dyakonov}}}\ and\ \bibinfo {author} {\bibfnamefont {V.~I.}\ \bibnamefont
  {{Perel}}},\ }\href {\doibase 10.1016/0375-9601(71)90196-4} {\bibfield
  {journal} {\bibinfo  {journal} {Phys. Lett. A}\ }\textbf {\bibinfo {volume}
  {35}},\ \bibinfo {pages} {459} (\bibinfo {year} {1971})}\BibitemShut
  {NoStop}%
\bibitem [{\citenamefont {Hirsch}(1999)}]{Hirsch1999}%
  \BibitemOpen
  \bibfield  {author} {\bibinfo {author} {\bibfnamefont {J.~E.}\ \bibnamefont
  {Hirsch}},\ }\href {\doibase 10.1103/PhysRevLett.83.1834} {\bibfield
  {journal} {\bibinfo  {journal} {Phys.\ Rev.\ Lett.}\ }\textbf {\bibinfo
  {volume} {83}},\ \bibinfo {pages} {1834} (\bibinfo {year}
  {1999})}\BibitemShut {NoStop}%
\bibitem [{\citenamefont {Sinova}\ \emph {et~al.}(2015)\citenamefont {Sinova},
  \citenamefont {Valenzuela}, \citenamefont {Wunderlich}, \citenamefont
  {Back},\ and\ \citenamefont {Jungwirth}}]{sinova2015Rev}%
  \BibitemOpen
  \bibfield  {author} {\bibinfo {author} {\bibfnamefont {J.}~\bibnamefont
  {Sinova}}, \bibinfo {author} {\bibfnamefont {S.~O.}\ \bibnamefont
  {Valenzuela}}, \bibinfo {author} {\bibfnamefont {J.}~\bibnamefont
  {Wunderlich}}, \bibinfo {author} {\bibfnamefont {C.~H.}\ \bibnamefont
  {Back}}, \ and\ \bibinfo {author} {\bibfnamefont {T.}~\bibnamefont
  {Jungwirth}},\ }\href {\doibase 10.1103/RevModPhys.87.1213} {\bibfield
  {journal} {\bibinfo  {journal} {Rev.\ Mod.\ Phys.}\ }\textbf {\bibinfo
  {volume} {87}},\ \bibinfo {pages} {1213} (\bibinfo {year}
  {2015})}\BibitemShut {NoStop}%
\bibitem [{\citenamefont {Manchon}\ \emph {et~al.}(2019)\citenamefont
  {Manchon}, \citenamefont {{\v{Z}}elezn{\`y}}, \citenamefont {Miron},
  \citenamefont {Jungwirth}, \citenamefont {Sinova}, \citenamefont {Thiaville},
  \citenamefont {Garello},\ and\ \citenamefont
  {Gambardella}}]{Manchon2019review}%
  \BibitemOpen
  \bibfield  {author} {\bibinfo {author} {\bibfnamefont {A.}~\bibnamefont
  {Manchon}}, \bibinfo {author} {\bibfnamefont {J.}~\bibnamefont
  {{\v{Z}}elezn{\`y}}}, \bibinfo {author} {\bibfnamefont {I.~M.}\ \bibnamefont
  {Miron}}, \bibinfo {author} {\bibfnamefont {T.}~\bibnamefont {Jungwirth}},
  \bibinfo {author} {\bibfnamefont {J.}~\bibnamefont {Sinova}}, \bibinfo
  {author} {\bibfnamefont {A.}~\bibnamefont {Thiaville}}, \bibinfo {author}
  {\bibfnamefont {K.}~\bibnamefont {Garello}}, \ and\ \bibinfo {author}
  {\bibfnamefont {P.}~\bibnamefont {Gambardella}},\ }\href
  {https://journals.aps.org/rmp/abstract/10.1103/RevModPhys.91.035004}
  {\bibfield  {journal} {\bibinfo  {journal} {Rev.\ Mod.\ Phys.}\ }\textbf
  {\bibinfo {volume} {91}},\ \bibinfo {pages} {035004} (\bibinfo {year}
  {2019})}\BibitemShut {NoStop}%
\bibitem [{\citenamefont {Bychkov}\ and\ \citenamefont
  {Rashba}(1984)}]{bychkov1984properties}%
  \BibitemOpen
  \bibfield  {author} {\bibinfo {author} {\bibfnamefont {Y.~A.}\ \bibnamefont
  {Bychkov}}\ and\ \bibinfo {author} {\bibfnamefont {{\'E}.~I.}\ \bibnamefont
  {Rashba}},\ }\href {http://www.jetp.ac.ru/cgi-bin/dn/e_071_02_0401.pdf}
  {\bibfield  {journal} {\bibinfo  {journal} {JETP Lett.}\ }\textbf {\bibinfo
  {volume} {39}},\ \bibinfo {pages} {78} (\bibinfo {year} {1984})}\BibitemShut
  {NoStop}%
\bibitem [{\citenamefont {Edelstein}(1990)}]{edelstein1990spin}%
  \BibitemOpen
  \bibfield  {author} {\bibinfo {author} {\bibfnamefont {V.~M.}\ \bibnamefont
  {Edelstein}},\ }\href {\doibase https://doi.org/10.1016/0038-1098(90)90963-C}
  {\bibfield  {journal} {\bibinfo  {journal} {Solid State Commun.}\ }\textbf
  {\bibinfo {volume} {73}},\ \bibinfo {pages} {233} (\bibinfo {year}
  {1990})}\BibitemShut {NoStop}%
\bibitem [{\citenamefont {{MacNeill}}\ \emph {et~al.}(2017)\citenamefont
  {{MacNeill}}, \citenamefont {{Stiehl}}, \citenamefont {{Guimaraes}},
  \citenamefont {{Buhrman}}, \citenamefont {{Park}},\ and\ \citenamefont
  {{Ralph}}}]{Macneil2017NP_WTe2}%
  \BibitemOpen
  \bibfield  {author} {\bibinfo {author} {\bibfnamefont {D.}~\bibnamefont
  {{MacNeill}}}, \bibinfo {author} {\bibfnamefont {G.~M.}\ \bibnamefont
  {{Stiehl}}}, \bibinfo {author} {\bibfnamefont {M.~H.~D.}\ \bibnamefont
  {{Guimaraes}}}, \bibinfo {author} {\bibfnamefont {R.~A.}\ \bibnamefont
  {{Buhrman}}}, \bibinfo {author} {\bibfnamefont {J.}~\bibnamefont {{Park}}}, \
  and\ \bibinfo {author} {\bibfnamefont {D.~C.}\ \bibnamefont {{Ralph}}},\
  }\href {\doibase 10.1038/nphys3933} {\bibfield  {journal} {\bibinfo
  {journal} {Nat.\ Phys.}\ }\textbf {\bibinfo {volume} {13}},\ \bibinfo {pages}
  {300} (\bibinfo {year} {2017})}\BibitemShut {NoStop}%
\bibitem [{\citenamefont {{Xie}}\ \emph {et~al.}(2021)\citenamefont {{Xie}},
  \citenamefont {{Lin}}, \citenamefont {{Sarkar}}, \citenamefont {{Shu}},
  \citenamefont {{Chen}}, \citenamefont {{Liu}}, \citenamefont {{Zhao}},
  \citenamefont {{Zhou}}, \citenamefont {{Wang}}, \citenamefont {{Zhou}},
  \citenamefont {{Grade{\v{c}}ak}},\ and\ \citenamefont
  {{Chen}}}]{xie2021field}%
  \BibitemOpen
  \bibfield  {author} {\bibinfo {author} {\bibfnamefont {Q.}~\bibnamefont
  {{Xie}}}, \bibinfo {author} {\bibfnamefont {W.}~\bibnamefont {{Lin}}},
  \bibinfo {author} {\bibfnamefont {S.}~\bibnamefont {{Sarkar}}}, \bibinfo
  {author} {\bibfnamefont {X.}~\bibnamefont {{Shu}}}, \bibinfo {author}
  {\bibfnamefont {S.}~\bibnamefont {{Chen}}}, \bibinfo {author} {\bibfnamefont
  {L.}~\bibnamefont {{Liu}}}, \bibinfo {author} {\bibfnamefont
  {T.}~\bibnamefont {{Zhao}}}, \bibinfo {author} {\bibfnamefont
  {C.}~\bibnamefont {{Zhou}}}, \bibinfo {author} {\bibfnamefont
  {H.}~\bibnamefont {{Wang}}}, \bibinfo {author} {\bibfnamefont
  {J.}~\bibnamefont {{Zhou}}}, \bibinfo {author} {\bibfnamefont
  {S.}~\bibnamefont {{Grade{\v{c}}ak}}}, \ and\ \bibinfo {author}
  {\bibfnamefont {J.}~\bibnamefont {{Chen}}},\ }\href
  {https://aip.scitation.org/doi/10.1063/5.0048926} {\bibfield  {journal}
  {\bibinfo  {journal} {APL Mater.}\ }\textbf {\bibinfo {volume} {9}},\
  \bibinfo {pages} {051114} (\bibinfo {year} {2021})}\BibitemShut {NoStop}%
\bibitem [{\citenamefont {{Nan}}\ \emph {et~al.}(2020)\citenamefont {{Nan}},
  \citenamefont {{Quintela}}, \citenamefont {{Irwin}}, \citenamefont
  {{Gurung}}, \citenamefont {{Shao}}, \citenamefont {{Gibbons}}, \citenamefont
  {{Campbell}}, \citenamefont {{Song}}, \citenamefont {{Choi}}, \citenamefont
  {{Guo}}, \citenamefont {{Johnson}}, \citenamefont {{Manuel}}, \citenamefont
  {{Chopdekar}}, \citenamefont {{Hallsteinsen}}, \citenamefont {{Tybell}},
  \citenamefont {{Ryan}}, \citenamefont {{Kim}}, \citenamefont {{Choi}},
  \citenamefont {{Radaelli}}, \citenamefont {{Ralph}}, \citenamefont
  {{Tsymbal}}, \citenamefont {{Rzchowski}},\ and\ \citenamefont
  {{Eom}}}]{nan2020controlling}%
  \BibitemOpen
  \bibfield  {author} {\bibinfo {author} {\bibfnamefont {T.}~\bibnamefont
  {{Nan}}}, \bibinfo {author} {\bibfnamefont {C.~X.}\ \bibnamefont
  {{Quintela}}}, \bibinfo {author} {\bibfnamefont {J.}~\bibnamefont {{Irwin}}},
  \bibinfo {author} {\bibfnamefont {G.}~\bibnamefont {{Gurung}}}, \bibinfo
  {author} {\bibfnamefont {D.~F.}\ \bibnamefont {{Shao}}}, \bibinfo {author}
  {\bibfnamefont {J.}~\bibnamefont {{Gibbons}}}, \bibinfo {author}
  {\bibfnamefont {N.}~\bibnamefont {{Campbell}}}, \bibinfo {author}
  {\bibfnamefont {K.}~\bibnamefont {{Song}}}, \bibinfo {author} {\bibfnamefont
  {S.~Y.}\ \bibnamefont {{Choi}}}, \bibinfo {author} {\bibfnamefont
  {L.}~\bibnamefont {{Guo}}}, \bibinfo {author} {\bibfnamefont {R.~D.}\
  \bibnamefont {{Johnson}}}, \bibinfo {author} {\bibfnamefont {P.}~\bibnamefont
  {{Manuel}}}, \bibinfo {author} {\bibfnamefont {R.~V.}\ \bibnamefont
  {{Chopdekar}}}, \bibinfo {author} {\bibfnamefont {I.}~\bibnamefont
  {{Hallsteinsen}}}, \bibinfo {author} {\bibfnamefont {T.}~\bibnamefont
  {{Tybell}}}, \bibinfo {author} {\bibfnamefont {P.~J.}\ \bibnamefont
  {{Ryan}}}, \bibinfo {author} {\bibfnamefont {J.~W.}\ \bibnamefont {{Kim}}},
  \bibinfo {author} {\bibfnamefont {Y.}~\bibnamefont {{Choi}}}, \bibinfo
  {author} {\bibfnamefont {P.~G.}\ \bibnamefont {{Radaelli}}}, \bibinfo
  {author} {\bibfnamefont {D.~C.}\ \bibnamefont {{Ralph}}}, \bibinfo {author}
  {\bibfnamefont {E.~Y.}\ \bibnamefont {{Tsymbal}}}, \bibinfo {author}
  {\bibfnamefont {M.~S.}\ \bibnamefont {{Rzchowski}}}, \ and\ \bibinfo {author}
  {\bibfnamefont {C.~B.}\ \bibnamefont {{Eom}}},\ }\href
  {https://www.nature.com/articles/s41467-020-17999-4} {\bibfield  {journal}
  {\bibinfo  {journal} {Nat.\ Commun.}\ }\textbf {\bibinfo {volume} {11}},\
  \bibinfo {pages} {1} (\bibinfo {year} {2020})}\BibitemShut {NoStop}%
\bibitem [{\citenamefont {You}\ \emph {et~al.}(2021)\citenamefont {You},
  \citenamefont {Bai}, \citenamefont {Feng}, \citenamefont {Fan}, \citenamefont
  {Han}, \citenamefont {Zhou}, \citenamefont {Zhou}, \citenamefont {Zhang},
  \citenamefont {Chen}, \citenamefont {Pan},\ and\ \citenamefont
  {Song}}]{you2021cluster}%
  \BibitemOpen
  \bibfield  {author} {\bibinfo {author} {\bibfnamefont {Y.}~\bibnamefont
  {You}}, \bibinfo {author} {\bibfnamefont {H.}~\bibnamefont {Bai}}, \bibinfo
  {author} {\bibfnamefont {X.}~\bibnamefont {Feng}}, \bibinfo {author}
  {\bibfnamefont {X.}~\bibnamefont {Fan}}, \bibinfo {author} {\bibfnamefont
  {L.}~\bibnamefont {Han}}, \bibinfo {author} {\bibfnamefont {X.}~\bibnamefont
  {Zhou}}, \bibinfo {author} {\bibfnamefont {Y.}~\bibnamefont {Zhou}}, \bibinfo
  {author} {\bibfnamefont {R.}~\bibnamefont {Zhang}}, \bibinfo {author}
  {\bibfnamefont {T.}~\bibnamefont {Chen}}, \bibinfo {author} {\bibfnamefont
  {F.}~\bibnamefont {Pan}}, \ and\ \bibinfo {author} {\bibfnamefont
  {C.}~\bibnamefont {Song}},\ }\href
  {https://www.nature.com/articles/s41467-021-26893-6} {\bibfield  {journal}
  {\bibinfo  {journal} {Nat. Commun.}\ }\textbf {\bibinfo {volume} {12}},\
  \bibinfo {pages} {1} (\bibinfo {year} {2021})}\BibitemShut {NoStop}%
\bibitem [{\citenamefont {Kondou}\ \emph {et~al.}(2021)\citenamefont {Kondou},
  \citenamefont {Chen}, \citenamefont {Tomita}, \citenamefont {Ikhlas},
  \citenamefont {Higo}, \citenamefont {MacDonald}, \citenamefont {Nakatsuji},\
  and\ \citenamefont {Otani}}]{kondou2021giant}%
  \BibitemOpen
  \bibfield  {author} {\bibinfo {author} {\bibfnamefont {K.}~\bibnamefont
  {Kondou}}, \bibinfo {author} {\bibfnamefont {H.}~\bibnamefont {Chen}},
  \bibinfo {author} {\bibfnamefont {T.}~\bibnamefont {Tomita}}, \bibinfo
  {author} {\bibfnamefont {M.}~\bibnamefont {Ikhlas}}, \bibinfo {author}
  {\bibfnamefont {T.}~\bibnamefont {Higo}}, \bibinfo {author} {\bibfnamefont
  {A.~H.}\ \bibnamefont {MacDonald}}, \bibinfo {author} {\bibfnamefont
  {S.}~\bibnamefont {Nakatsuji}}, \ and\ \bibinfo {author} {\bibfnamefont
  {Y.}~\bibnamefont {Otani}},\ }\href
  {https://www.nature.com/articles/s41467-021-26453-y} {\bibfield  {journal}
  {\bibinfo  {journal} {Nat. Commun.}\ }\textbf {\bibinfo {volume} {12}},\
  \bibinfo {pages} {1} (\bibinfo {year} {2021})}\BibitemShut {NoStop}%
\bibitem [{\citenamefont {{Zhou}}\ \emph {et~al.}(2020)\citenamefont {{Zhou}},
  \citenamefont {{Shu}}, \citenamefont {{Liu}}, \citenamefont {{Wang}},
  \citenamefont {{Lin}}, \citenamefont {{Chen}}, \citenamefont {{Liu}},
  \citenamefont {{Xie}}, \citenamefont {{Hong}}, \citenamefont {{Yang}},
  \citenamefont {{Yan}}, \citenamefont {{Han}},\ and\ \citenamefont
  {{Chen}}}]{zhou2020magnetic}%
  \BibitemOpen
  \bibfield  {author} {\bibinfo {author} {\bibfnamefont {J.}~\bibnamefont
  {{Zhou}}}, \bibinfo {author} {\bibfnamefont {X.}~\bibnamefont {{Shu}}},
  \bibinfo {author} {\bibfnamefont {Y.}~\bibnamefont {{Liu}}}, \bibinfo
  {author} {\bibfnamefont {X.}~\bibnamefont {{Wang}}}, \bibinfo {author}
  {\bibfnamefont {W.}~\bibnamefont {{Lin}}}, \bibinfo {author} {\bibfnamefont
  {S.}~\bibnamefont {{Chen}}}, \bibinfo {author} {\bibfnamefont
  {L.}~\bibnamefont {{Liu}}}, \bibinfo {author} {\bibfnamefont
  {Q.}~\bibnamefont {{Xie}}}, \bibinfo {author} {\bibfnamefont
  {T.}~\bibnamefont {{Hong}}}, \bibinfo {author} {\bibfnamefont
  {P.}~\bibnamefont {{Yang}}}, \bibinfo {author} {\bibfnamefont
  {B.}~\bibnamefont {{Yan}}}, \bibinfo {author} {\bibfnamefont
  {X.}~\bibnamefont {{Han}}}, \ and\ \bibinfo {author} {\bibfnamefont
  {J.}~\bibnamefont {{Chen}}},\ }\href
  {https://journals.aps.org/prb/abstract/10.1103/PhysRevB.101.184403}
  {\bibfield  {journal} {\bibinfo  {journal} {Phys.\ Rev.\ B}\ }\textbf
  {\bibinfo {volume} {101}},\ \bibinfo {pages} {184403} (\bibinfo {year}
  {2020})}\BibitemShut {NoStop}%
\bibitem [{\citenamefont {Shindou}\ and\ \citenamefont
  {Nagaosa}(2001)}]{shindou2001orbital}%
  \BibitemOpen
  \bibfield  {author} {\bibinfo {author} {\bibfnamefont {R.}~\bibnamefont
  {Shindou}}\ and\ \bibinfo {author} {\bibfnamefont {N.}~\bibnamefont
  {Nagaosa}},\ }\href
  {https://journals.aps.org/prl/abstract/10.1103/PhysRevLett.87.116801}
  {\bibfield  {journal} {\bibinfo  {journal} {Phys.\ Rev.\ Lett.}\ }\textbf
  {\bibinfo {volume} {87}},\ \bibinfo {pages} {116801} (\bibinfo {year}
  {2001})}\BibitemShut {NoStop}%
\bibitem [{\citenamefont {K{\"u}bler}\ and\ \citenamefont
  {Felser}(2014)}]{kubler2014non}%
  \BibitemOpen
  \bibfield  {author} {\bibinfo {author} {\bibfnamefont {J.}~\bibnamefont
  {K{\"u}bler}}\ and\ \bibinfo {author} {\bibfnamefont {C.}~\bibnamefont
  {Felser}},\ }\href
  {https://iopscience.iop.org/article/10.1209/0295-5075/108/67001/meta}
  {\bibfield  {journal} {\bibinfo  {journal} {EPL}\ }\textbf {\bibinfo {volume}
  {108}},\ \bibinfo {pages} {67001} (\bibinfo {year} {2014})}\BibitemShut
  {NoStop}%
\bibitem [{\citenamefont {Chen}\ \emph {et~al.}(2014)\citenamefont {Chen},
  \citenamefont {Niu},\ and\ \citenamefont {MacDonald}}]{chen2014anomalous}%
  \BibitemOpen
  \bibfield  {author} {\bibinfo {author} {\bibfnamefont {H.}~\bibnamefont
  {Chen}}, \bibinfo {author} {\bibfnamefont {Q.}~\bibnamefont {Niu}}, \ and\
  \bibinfo {author} {\bibfnamefont {A.~H.}\ \bibnamefont {MacDonald}},\ }\href
  {http://dx.doi.org/10.1103/PhysRevLett.112.017205} {\bibfield  {journal}
  {\bibinfo  {journal} {Phys.\ Rev.\ Lett.}\ }\textbf {\bibinfo {volume}
  {112}},\ \bibinfo {pages} {017205} (\bibinfo {year} {2014})}\BibitemShut
  {NoStop}%
\bibitem [{\citenamefont {Nagaosa}\ \emph {et~al.}(2010)\citenamefont
  {Nagaosa}, \citenamefont {Sinova}, \citenamefont {Onoda}, \citenamefont
  {MacDonald},\ and\ \citenamefont {Ong}}]{nagaosa2010anomalous}%
  \BibitemOpen
  \bibfield  {author} {\bibinfo {author} {\bibfnamefont {N.}~\bibnamefont
  {Nagaosa}}, \bibinfo {author} {\bibfnamefont {J.}~\bibnamefont {Sinova}},
  \bibinfo {author} {\bibfnamefont {S.}~\bibnamefont {Onoda}}, \bibinfo
  {author} {\bibfnamefont {A.~H.}\ \bibnamefont {MacDonald}}, \ and\ \bibinfo
  {author} {\bibfnamefont {N.~P.}\ \bibnamefont {Ong}},\ }\href
  {https://journals.aps.org/rmp/abstract/10.1103/RevModPhys.82.1539} {\bibfield
   {journal} {\bibinfo  {journal} {Rev. Mod. Phys.}\ }\textbf {\bibinfo
  {volume} {82}},\ \bibinfo {pages} {1539} (\bibinfo {year}
  {2010})}\BibitemShut {NoStop}%
\bibitem [{\citenamefont {Higo}\ \emph {et~al.}(2018)\citenamefont {Higo},
  \citenamefont {Qu}, \citenamefont {Li}, \citenamefont {Chien}, \citenamefont
  {Otani},\ and\ \citenamefont {Nakatsuji}}]{higo2018anomalous}%
  \BibitemOpen
  \bibfield  {author} {\bibinfo {author} {\bibfnamefont {T.}~\bibnamefont
  {Higo}}, \bibinfo {author} {\bibfnamefont {D.}~\bibnamefont {Qu}}, \bibinfo
  {author} {\bibfnamefont {Y.}~\bibnamefont {Li}}, \bibinfo {author}
  {\bibfnamefont {C.}~\bibnamefont {Chien}}, \bibinfo {author} {\bibfnamefont
  {Y.}~\bibnamefont {Otani}}, \ and\ \bibinfo {author} {\bibfnamefont
  {S.}~\bibnamefont {Nakatsuji}},\ }\href {https://doi.org/10.1063/1.5064697}
  {\bibfield  {journal} {\bibinfo  {journal} {Appl.\ Phys.\ Lett.}\ }\textbf
  {\bibinfo {volume} {113}},\ \bibinfo {pages} {202402} (\bibinfo {year}
  {2018})}\BibitemShut {NoStop}%
\bibitem [{\citenamefont {Han}\ \emph {et~al.}(2018)\citenamefont {Han},
  \citenamefont {Otani},\ and\ \citenamefont {Maekawa}}]{han2018quantum}%
  \BibitemOpen
  \bibfield  {author} {\bibinfo {author} {\bibfnamefont {W.}~\bibnamefont
  {Han}}, \bibinfo {author} {\bibfnamefont {Y.}~\bibnamefont {Otani}}, \ and\
  \bibinfo {author} {\bibfnamefont {S.}~\bibnamefont {Maekawa}},\ }\href
  {https://www.nature.com/articles/s41535-018-0100-9} {\bibfield  {journal}
  {\bibinfo  {journal} {npj Quantum Mater.}\ }\textbf {\bibinfo {volume} {3}},\
  \bibinfo {pages} {1} (\bibinfo {year} {2018})}\BibitemShut {NoStop}%
\bibitem [{\citenamefont {Nogu{\'e}s}\ and\ \citenamefont
  {Schuller}(1999)}]{nogues1999exchange}%
  \BibitemOpen
  \bibfield  {author} {\bibinfo {author} {\bibfnamefont {J.}~\bibnamefont
  {Nogu{\'e}s}}\ and\ \bibinfo {author} {\bibfnamefont {I.~K.}\ \bibnamefont
  {Schuller}},\ }\href
  {https://www.sciencedirect.com/science/article/pii/S0304885398002662}
  {\bibfield  {journal} {\bibinfo  {journal} {J.\ Magn.\ Magn.\ Mater.}\
  }\textbf {\bibinfo {volume} {192}},\ \bibinfo {pages} {203} (\bibinfo {year}
  {1999})}\BibitemShut {NoStop}%
\bibitem [{\citenamefont {{Oh}}\ \emph {et~al.}(2016)\citenamefont {{Oh}},
  \citenamefont {{Chris Baek}}, \citenamefont {{Kim}}, \citenamefont {{Lee}},
  \citenamefont {{Lee}}, \citenamefont {{Yang}}, \citenamefont {{Park}},
  \citenamefont {{Lee}}, \citenamefont {{Kim}}, \citenamefont {{Go}},
  \citenamefont {{Jeong}}, \citenamefont {{Min}}, \citenamefont {{Lee}},
  \citenamefont {{Lee}},\ and\ \citenamefont {{Park}}}]{oh2016field}%
  \BibitemOpen
  \bibfield  {author} {\bibinfo {author} {\bibfnamefont {Y.-W.}\ \bibnamefont
  {{Oh}}}, \bibinfo {author} {\bibfnamefont {S.-H.}\ \bibnamefont {{Chris
  Baek}}}, \bibinfo {author} {\bibfnamefont {Y.~M.}\ \bibnamefont {{Kim}}},
  \bibinfo {author} {\bibfnamefont {H.~Y.}\ \bibnamefont {{Lee}}}, \bibinfo
  {author} {\bibfnamefont {K.-D.}\ \bibnamefont {{Lee}}}, \bibinfo {author}
  {\bibfnamefont {C.-G.}\ \bibnamefont {{Yang}}}, \bibinfo {author}
  {\bibfnamefont {E.-S.}\ \bibnamefont {{Park}}}, \bibinfo {author}
  {\bibfnamefont {K.-S.}\ \bibnamefont {{Lee}}}, \bibinfo {author}
  {\bibfnamefont {K.-W.}\ \bibnamefont {{Kim}}}, \bibinfo {author}
  {\bibfnamefont {G.}~\bibnamefont {{Go}}}, \bibinfo {author} {\bibfnamefont
  {J.-R.}\ \bibnamefont {{Jeong}}}, \bibinfo {author} {\bibfnamefont {B.-C.}\
  \bibnamefont {{Min}}}, \bibinfo {author} {\bibfnamefont {H.-W.}\ \bibnamefont
  {{Lee}}}, \bibinfo {author} {\bibfnamefont {K.-J.}\ \bibnamefont {{Lee}}}, \
  and\ \bibinfo {author} {\bibfnamefont {B.-G.}\ \bibnamefont {{Park}}},\
  }\href {https://www.nature.com/articles/nnano.2016.109} {\bibfield  {journal}
  {\bibinfo  {journal} {Nat.\ Nano.}\ }\textbf {\bibinfo {volume} {11}},\
  \bibinfo {pages} {878} (\bibinfo {year} {2016})}\BibitemShut {NoStop}%
\bibitem [{\citenamefont {Fukami}\ \emph {et~al.}(2016)\citenamefont {Fukami},
  \citenamefont {Zhang}, \citenamefont {DuttaGupta}, \citenamefont {Kurenkov},\
  and\ \citenamefont {Ohno}}]{fukami2016magnetization}%
  \BibitemOpen
  \bibfield  {author} {\bibinfo {author} {\bibfnamefont {S.}~\bibnamefont
  {Fukami}}, \bibinfo {author} {\bibfnamefont {C.}~\bibnamefont {Zhang}},
  \bibinfo {author} {\bibfnamefont {S.}~\bibnamefont {DuttaGupta}}, \bibinfo
  {author} {\bibfnamefont {A.}~\bibnamefont {Kurenkov}}, \ and\ \bibinfo
  {author} {\bibfnamefont {H.}~\bibnamefont {Ohno}},\ }\href
  {https://www.nature.com/articles/nmat4566} {\bibfield  {journal} {\bibinfo
  {journal} {Nat.\ Mater.}\ }\textbf {\bibinfo {volume} {15}},\ \bibinfo
  {pages} {535} (\bibinfo {year} {2016})}\BibitemShut {NoStop}%
\bibitem [{\citenamefont {van~den Brink}\ \emph {et~al.}(2016)\citenamefont
  {van~den Brink}, \citenamefont {Vermijs}, \citenamefont {Solignac},
  \citenamefont {Koo}, \citenamefont {Kohlhepp}, \citenamefont {Swagten},\ and\
  \citenamefont {Koopmans}}]{van2016field}%
  \BibitemOpen
  \bibfield  {author} {\bibinfo {author} {\bibfnamefont {A.}~\bibnamefont
  {van~den Brink}}, \bibinfo {author} {\bibfnamefont {G.}~\bibnamefont
  {Vermijs}}, \bibinfo {author} {\bibfnamefont {A.}~\bibnamefont {Solignac}},
  \bibinfo {author} {\bibfnamefont {J.}~\bibnamefont {Koo}}, \bibinfo {author}
  {\bibfnamefont {J.~T.}\ \bibnamefont {Kohlhepp}}, \bibinfo {author}
  {\bibfnamefont {H.~J.}\ \bibnamefont {Swagten}}, \ and\ \bibinfo {author}
  {\bibfnamefont {B.}~\bibnamefont {Koopmans}},\ }\href
  {https://www.nature.com/articles/ncomms10854} {\bibfield  {journal} {\bibinfo
   {journal} {Nat.\ Commun.}\ }\textbf {\bibinfo {volume} {7}},\ \bibinfo
  {pages} {1} (\bibinfo {year} {2016})}\BibitemShut {NoStop}%
\bibitem [{\citenamefont {Migliorini}\ \emph {et~al.}(2018)\citenamefont
  {Migliorini}, \citenamefont {Kuerbanjiang}, \citenamefont {Huminiuc},
  \citenamefont {Kepaptsoglou}, \citenamefont {Mu{\~n}oz}, \citenamefont
  {Cu{\~n}ado}, \citenamefont {Camarero}, \citenamefont {Aroca}, \citenamefont
  {Vallejo-Fern{\'a}ndez}, \citenamefont {Lazarov},\ and\ \citenamefont
  {Prieto}}]{migliorini2018spontaneous}%
  \BibitemOpen
  \bibfield  {author} {\bibinfo {author} {\bibfnamefont {A.}~\bibnamefont
  {Migliorini}}, \bibinfo {author} {\bibfnamefont {B.}~\bibnamefont
  {Kuerbanjiang}}, \bibinfo {author} {\bibfnamefont {T.}~\bibnamefont
  {Huminiuc}}, \bibinfo {author} {\bibfnamefont {D.}~\bibnamefont
  {Kepaptsoglou}}, \bibinfo {author} {\bibfnamefont {M.}~\bibnamefont
  {Mu{\~n}oz}}, \bibinfo {author} {\bibfnamefont {J.}~\bibnamefont
  {Cu{\~n}ado}}, \bibinfo {author} {\bibfnamefont {J.}~\bibnamefont
  {Camarero}}, \bibinfo {author} {\bibfnamefont {C.}~\bibnamefont {Aroca}},
  \bibinfo {author} {\bibfnamefont {G.}~\bibnamefont {Vallejo-Fern{\'a}ndez}},
  \bibinfo {author} {\bibfnamefont {V.}~\bibnamefont {Lazarov}}, \ and\
  \bibinfo {author} {\bibfnamefont {J.}~\bibnamefont {Prieto}},\ }\href
  {https://www.nature.com/articles/nmat5030} {\bibfield  {journal} {\bibinfo
  {journal} {Nat.\ Mater.}\ }\textbf {\bibinfo {volume} {17}},\ \bibinfo
  {pages} {28} (\bibinfo {year} {2018})}\BibitemShut {NoStop}%
\bibitem [{\citenamefont {Tsunoda}\ \emph {et~al.}(2006)\citenamefont
  {Tsunoda}, \citenamefont {Imakita}, \citenamefont {Naka},\ and\ \citenamefont
  {Takahashi}}]{tsunoda2006l12}%
  \BibitemOpen
  \bibfield  {author} {\bibinfo {author} {\bibfnamefont {M.}~\bibnamefont
  {Tsunoda}}, \bibinfo {author} {\bibfnamefont {K.-i.}\ \bibnamefont
  {Imakita}}, \bibinfo {author} {\bibfnamefont {M.}~\bibnamefont {Naka}}, \
  and\ \bibinfo {author} {\bibfnamefont {M.}~\bibnamefont {Takahashi}},\ }\href
  {https://www.sciencedirect.com/science/article/pii/S0304885306001739}
  {\bibfield  {journal} {\bibinfo  {journal} {J.\ Magn.\ Magn.\ Mater.}\
  }\textbf {\bibinfo {volume} {304}},\ \bibinfo {pages} {55} (\bibinfo {year}
  {2006})}\BibitemShut {NoStop}%
\bibitem [{\citenamefont {Parratt}(1954)}]{parratt1954surface}%
  \BibitemOpen
  \bibfield  {author} {\bibinfo {author} {\bibfnamefont {L.~G.}\ \bibnamefont
  {Parratt}},\ }\href
  {https://journals.aps.org/pr/abstract/10.1103/PhysRev.95.359} {\bibfield
  {journal} {\bibinfo  {journal} {Phys. Rev.}\ }\textbf {\bibinfo {volume}
  {95}},\ \bibinfo {pages} {359} (\bibinfo {year} {1954})}\BibitemShut
  {NoStop}%
\bibitem [{\citenamefont {Bansal}\ \emph {et~al.}(2017)\citenamefont {Bansal},
  \citenamefont {Behera}, \citenamefont {Kumar},\ and\ \citenamefont
  {Muduli}}]{bansal2017crystalline}%
  \BibitemOpen
  \bibfield  {author} {\bibinfo {author} {\bibfnamefont {R.}~\bibnamefont
  {Bansal}}, \bibinfo {author} {\bibfnamefont {N.}~\bibnamefont {Behera}},
  \bibinfo {author} {\bibfnamefont {A.}~\bibnamefont {Kumar}}, \ and\ \bibinfo
  {author} {\bibfnamefont {P.}~\bibnamefont {Muduli}},\ }\href
  {https://aip.scitation.org/doi/pdf/10.1063/1.4983677} {\bibfield  {journal}
  {\bibinfo  {journal} {Appl.\ Phys.\ Lett.}\ }\textbf {\bibinfo {volume}
  {110}},\ \bibinfo {pages} {202402} (\bibinfo {year} {2017})}\BibitemShut
  {NoStop}%
\bibitem [{\citenamefont {Khomenko}\ \emph {et~al.}(2008)\citenamefont
  {Khomenko}, \citenamefont {Chechenin}, \citenamefont {Goikhman},\ and\
  \citenamefont {Zenkevich}}]{khomenko2008exchange}%
  \BibitemOpen
  \bibfield  {author} {\bibinfo {author} {\bibfnamefont {E.~V.}\ \bibnamefont
  {Khomenko}}, \bibinfo {author} {\bibfnamefont {N.~G.}\ \bibnamefont
  {Chechenin}}, \bibinfo {author} {\bibfnamefont {A.~Y.}\ \bibnamefont
  {Goikhman}}, \ and\ \bibinfo {author} {\bibfnamefont {A.~V.}\ \bibnamefont
  {Zenkevich}},\ }\href
  {https://link.springer.com/article/10.1134/S0021364008210121} {\bibfield
  {journal} {\bibinfo  {journal} {JETP Lett.}\ }\textbf {\bibinfo {volume}
  {88}},\ \bibinfo {pages} {602} (\bibinfo {year} {2008})}\BibitemShut
  {NoStop}%
\bibitem [{\citenamefont {Kumar}\ \emph {et~al.}(2021)\citenamefont {Kumar},
  \citenamefont {Sharma}, \citenamefont {Ali~Khan}, \citenamefont {Murapaka},
  \citenamefont {Lim}, \citenamefont {Lew}, \citenamefont {Chaudhary},\ and\
  \citenamefont {Muduli}}]{kumar2021large}%
  \BibitemOpen
  \bibfield  {author} {\bibinfo {author} {\bibfnamefont {A.}~\bibnamefont
  {Kumar}}, \bibinfo {author} {\bibfnamefont {R.}~\bibnamefont {Sharma}},
  \bibinfo {author} {\bibfnamefont {K.~I.}\ \bibnamefont {Ali~Khan}}, \bibinfo
  {author} {\bibfnamefont {C.}~\bibnamefont {Murapaka}}, \bibinfo {author}
  {\bibfnamefont {G.~J.}\ \bibnamefont {Lim}}, \bibinfo {author} {\bibfnamefont
  {W.~S.}\ \bibnamefont {Lew}}, \bibinfo {author} {\bibfnamefont
  {S.}~\bibnamefont {Chaudhary}}, \ and\ \bibinfo {author} {\bibfnamefont
  {P.~K.}\ \bibnamefont {Muduli}},\ }\href
  {https://doi.org/10.1021/acsaelm.1c00361} {\bibfield  {journal} {\bibinfo
  {journal} {ACS Appl. Electron. Mater.}\ }\textbf {\bibinfo {volume} {3}},\
  \bibinfo {pages} {3139} (\bibinfo {year} {2021})}\BibitemShut {NoStop}%
\bibitem [{\citenamefont {Liu}\ \emph {et~al.}(2011)\citenamefont {Liu},
  \citenamefont {Moriyama}, \citenamefont {Ralph},\ and\ \citenamefont
  {Buhrman}}]{Liu2011}%
  \BibitemOpen
  \bibfield  {author} {\bibinfo {author} {\bibfnamefont {L.}~\bibnamefont
  {Liu}}, \bibinfo {author} {\bibfnamefont {T.}~\bibnamefont {Moriyama}},
  \bibinfo {author} {\bibfnamefont {D.~C.}\ \bibnamefont {Ralph}}, \ and\
  \bibinfo {author} {\bibfnamefont {R.~A.}\ \bibnamefont {Buhrman}},\ }\href
  {\doibase 10.1103/PhysRevLett.106.036601} {\bibfield  {journal} {\bibinfo
  {journal} {Phys. Rev. Lett.}\ }\textbf {\bibinfo {volume} {106}},\ \bibinfo
  {pages} {036601} (\bibinfo {year} {2011})}\BibitemShut {NoStop}%
\bibitem [{\citenamefont {Bose}\ \emph {et~al.}(2022)\citenamefont {Bose},
  \citenamefont {Schreiber}, \citenamefont {Jain}, \citenamefont {Shao},
  \citenamefont {Nair}, \citenamefont {Sun}, \citenamefont {Zhang},
  \citenamefont {Muller}, \citenamefont {Tsymbal}, \citenamefont {Schlom},\
  and\ \citenamefont {Ralph}}]{bose2022tilted}%
  \BibitemOpen
  \bibfield  {author} {\bibinfo {author} {\bibfnamefont {A.}~\bibnamefont
  {Bose}}, \bibinfo {author} {\bibfnamefont {N.~J.}\ \bibnamefont {Schreiber}},
  \bibinfo {author} {\bibfnamefont {R.}~\bibnamefont {Jain}}, \bibinfo {author}
  {\bibfnamefont {D.-F.}\ \bibnamefont {Shao}}, \bibinfo {author}
  {\bibfnamefont {H.~P.}\ \bibnamefont {Nair}}, \bibinfo {author}
  {\bibfnamefont {J.}~\bibnamefont {Sun}}, \bibinfo {author} {\bibfnamefont
  {X.~S.}\ \bibnamefont {Zhang}}, \bibinfo {author} {\bibfnamefont {D.~A.}\
  \bibnamefont {Muller}}, \bibinfo {author} {\bibfnamefont {E.~Y.}\
  \bibnamefont {Tsymbal}}, \bibinfo {author} {\bibfnamefont {D.~G.}\
  \bibnamefont {Schlom}}, \ and\ \bibinfo {author} {\bibfnamefont {D.~C.}\
  \bibnamefont {Ralph}},\ }\href
  {https://www.nature.com/articles/s41928-022-00744-8} {\bibfield  {journal}
  {\bibinfo  {journal} {Nat. Electron.}\ }\textbf {\bibinfo {volume} {5}},\
  \bibinfo {pages} {267} (\bibinfo {year} {2022})}\BibitemShut {NoStop}%
\bibitem [{\citenamefont {Kittel}(1948)}]{Kittle1948}%
  \BibitemOpen
  \bibfield  {author} {\bibinfo {author} {\bibfnamefont {C.}~\bibnamefont
  {Kittel}},\ }\href {\doibase 10.1103/PhysRev.73.155} {\bibfield  {journal}
  {\bibinfo  {journal} {Phys.\ Rev.}\ }\textbf {\bibinfo {volume} {73}},\
  \bibinfo {pages} {155} (\bibinfo {year} {1948})}\BibitemShut {NoStop}%
\bibitem [{\citenamefont {Tshitoyan}\ \emph {et~al.}(2015)\citenamefont
  {Tshitoyan}, \citenamefont {Ciccarelli}, \citenamefont {Mihai}, \citenamefont
  {Ali}, \citenamefont {Irvine}, \citenamefont {Moore}, \citenamefont
  {Jungwirth},\ and\ \citenamefont {Ferguson}}]{tshitoyan2015electrical}%
  \BibitemOpen
  \bibfield  {author} {\bibinfo {author} {\bibfnamefont {V.}~\bibnamefont
  {Tshitoyan}}, \bibinfo {author} {\bibfnamefont {C.}~\bibnamefont
  {Ciccarelli}}, \bibinfo {author} {\bibfnamefont {A.}~\bibnamefont {Mihai}},
  \bibinfo {author} {\bibfnamefont {M.}~\bibnamefont {Ali}}, \bibinfo {author}
  {\bibfnamefont {A.}~\bibnamefont {Irvine}}, \bibinfo {author} {\bibfnamefont
  {T.}~\bibnamefont {Moore}}, \bibinfo {author} {\bibfnamefont
  {T.}~\bibnamefont {Jungwirth}}, \ and\ \bibinfo {author} {\bibfnamefont
  {A.}~\bibnamefont {Ferguson}},\ }\href
  {https://journals.aps.org/prb/abstract/10.1103/PhysRevB.92.214406} {\bibfield
   {journal} {\bibinfo  {journal} {Phys.\ Rev.\ B}\ }\textbf {\bibinfo {volume}
  {92}},\ \bibinfo {pages} {214406} (\bibinfo {year} {2015})}\BibitemShut
  {NoStop}%
\bibitem [{\citenamefont {Amin}\ \emph {et~al.}(2020)\citenamefont {Amin},
  \citenamefont {Haney},\ and\ \citenamefont {Stiles}}]{amin2020interfacial}%
  \BibitemOpen
  \bibfield  {author} {\bibinfo {author} {\bibfnamefont {V.~P.}\ \bibnamefont
  {Amin}}, \bibinfo {author} {\bibfnamefont {P.~M.}\ \bibnamefont {Haney}}, \
  and\ \bibinfo {author} {\bibfnamefont {M.~D.}\ \bibnamefont {Stiles}},\
  }\href {https://aip.scitation.org/doi/10.1063/5.0024019} {\bibfield
  {journal} {\bibinfo  {journal} {J.\ Appl.\ Phys.}\ }\textbf {\bibinfo
  {volume} {128}},\ \bibinfo {pages} {151101} (\bibinfo {year}
  {2020})}\BibitemShut {NoStop}%
\bibitem [{\citenamefont {Lifshits}\ and\ \citenamefont
  {Dyakonov}(2009)}]{lifshits2009swapping}%
  \BibitemOpen
  \bibfield  {author} {\bibinfo {author} {\bibfnamefont {M.~B.}\ \bibnamefont
  {Lifshits}}\ and\ \bibinfo {author} {\bibfnamefont {M.~I.}\ \bibnamefont
  {Dyakonov}},\ }\href
  {https://journals.aps.org/prl/abstract/10.1103/PhysRevLett.103.186601}
  {\bibfield  {journal} {\bibinfo  {journal} {Phys.\ Rev.\ Lett.}\ }\textbf
  {\bibinfo {volume} {103}},\ \bibinfo {pages} {186601} (\bibinfo {year}
  {2009})}\BibitemShut {NoStop}%
\bibitem [{\citenamefont {Hazra}\ \emph {et~al.}(2022)\citenamefont {Hazra},
  \citenamefont {Pal}, \citenamefont {Jeon}, \citenamefont {Neumann},
  \citenamefont {Goebel}, \citenamefont {Grover}, \citenamefont {Deniz},
  \citenamefont {Styervoyedov}, \citenamefont {Meyerheim}, \citenamefont
  {Mertig}, \citenamefont {Yang},\ and\ \citenamefont
  {Parkin}}]{hazra2022generation}%
  \BibitemOpen
  \bibfield  {author} {\bibinfo {author} {\bibfnamefont {B.~K.}\ \bibnamefont
  {Hazra}}, \bibinfo {author} {\bibfnamefont {B.}~\bibnamefont {Pal}}, \bibinfo
  {author} {\bibfnamefont {J.-C.}\ \bibnamefont {Jeon}}, \bibinfo {author}
  {\bibfnamefont {R.~R.}\ \bibnamefont {Neumann}}, \bibinfo {author}
  {\bibfnamefont {B.}~\bibnamefont {Goebel}}, \bibinfo {author} {\bibfnamefont
  {B.}~\bibnamefont {Grover}}, \bibinfo {author} {\bibfnamefont
  {H.}~\bibnamefont {Deniz}}, \bibinfo {author} {\bibfnamefont
  {A.}~\bibnamefont {Styervoyedov}}, \bibinfo {author} {\bibfnamefont
  {H.}~\bibnamefont {Meyerheim}}, \bibinfo {author} {\bibfnamefont
  {I.}~\bibnamefont {Mertig}}, \bibinfo {author} {\bibfnamefont {S.-H.}\
  \bibnamefont {Yang}}, \ and\ \bibinfo {author} {\bibfnamefont {S.~S.~P.}\
  \bibnamefont {Parkin}},\ }\href {https://arxiv.org/abs/2211.12398} {\bibfield
   {journal} {\bibinfo  {journal} {arXiv:2211.12398}\ } (\bibinfo {year}
  {2022})}\BibitemShut {NoStop}%
\bibitem [{\citenamefont {Saglam}\ \emph {et~al.}(2018)\citenamefont {Saglam},
  \citenamefont {Rojas-Sanchez}, \citenamefont {Petit}, \citenamefont {Hehn},
  \citenamefont {Zhang}, \citenamefont {Pearson}, \citenamefont {Mangin},\ and\
  \citenamefont {Hoffmann}}]{saglam2018independence}%
  \BibitemOpen
  \bibfield  {author} {\bibinfo {author} {\bibfnamefont {H.}~\bibnamefont
  {Saglam}}, \bibinfo {author} {\bibfnamefont {J.~C.}\ \bibnamefont
  {Rojas-Sanchez}}, \bibinfo {author} {\bibfnamefont {S.}~\bibnamefont
  {Petit}}, \bibinfo {author} {\bibfnamefont {M.}~\bibnamefont {Hehn}},
  \bibinfo {author} {\bibfnamefont {W.}~\bibnamefont {Zhang}}, \bibinfo
  {author} {\bibfnamefont {J.~E.}\ \bibnamefont {Pearson}}, \bibinfo {author}
  {\bibfnamefont {S.}~\bibnamefont {Mangin}}, \ and\ \bibinfo {author}
  {\bibfnamefont {A.}~\bibnamefont {Hoffmann}},\ }\href
  {https://doi.org/10.1103/PhysRevB.98.094407} {\bibfield  {journal} {\bibinfo
  {journal} {Phys.\ Rev.\ B}\ }\textbf {\bibinfo {volume} {98}},\ \bibinfo
  {pages} {094407} (\bibinfo {year} {2018})}\BibitemShut {NoStop}%
\bibitem [{\citenamefont {{Zhou}}\ \emph {et~al.}(2019)\citenamefont {{Zhou}},
  \citenamefont {{Wang}}, \citenamefont {{Liu}}, \citenamefont {{Yu}},
  \citenamefont {{Fu}}, \citenamefont {{Liu}}, \citenamefont {{Chen}},
  \citenamefont {{Deng}}, \citenamefont {{Lin}}, \citenamefont {{Shu}},
  \citenamefont {{Yoong}}, \citenamefont {{Hong}}, \citenamefont {{Matsuda}},
  \citenamefont {{Yang}}, \citenamefont {{Adams}}, \citenamefont {{Yan}},
  \citenamefont {{Han}},\ and\ \citenamefont {{Chen}}}]{zhou2019large}%
  \BibitemOpen
  \bibfield  {author} {\bibinfo {author} {\bibfnamefont {J.}~\bibnamefont
  {{Zhou}}}, \bibinfo {author} {\bibfnamefont {X.}~\bibnamefont {{Wang}}},
  \bibinfo {author} {\bibfnamefont {Y.}~\bibnamefont {{Liu}}}, \bibinfo
  {author} {\bibfnamefont {J.}~\bibnamefont {{Yu}}}, \bibinfo {author}
  {\bibfnamefont {H.}~\bibnamefont {{Fu}}}, \bibinfo {author} {\bibfnamefont
  {L.}~\bibnamefont {{Liu}}}, \bibinfo {author} {\bibfnamefont
  {S.}~\bibnamefont {{Chen}}}, \bibinfo {author} {\bibfnamefont
  {J.}~\bibnamefont {{Deng}}}, \bibinfo {author} {\bibfnamefont
  {W.}~\bibnamefont {{Lin}}}, \bibinfo {author} {\bibfnamefont
  {X.}~\bibnamefont {{Shu}}}, \bibinfo {author} {\bibfnamefont {H.~Y.}\
  \bibnamefont {{Yoong}}}, \bibinfo {author} {\bibfnamefont {T.}~\bibnamefont
  {{Hong}}}, \bibinfo {author} {\bibfnamefont {M.}~\bibnamefont {{Matsuda}}},
  \bibinfo {author} {\bibfnamefont {P.}~\bibnamefont {{Yang}}}, \bibinfo
  {author} {\bibfnamefont {S.}~\bibnamefont {{Adams}}}, \bibinfo {author}
  {\bibfnamefont {B.}~\bibnamefont {{Yan}}}, \bibinfo {author} {\bibfnamefont
  {X.}~\bibnamefont {{Han}}}, \ and\ \bibinfo {author} {\bibfnamefont
  {J.}~\bibnamefont {{Chen}}},\ }\href
  {https://ui.adsabs.harvard.edu/abs/2019SciA....5.6696Z} {\bibfield  {journal}
  {\bibinfo  {journal} {Sci. Adv.}\ }\textbf {\bibinfo {volume} {5}},\ \bibinfo
  {pages} {eaau6696} (\bibinfo {year} {2019})}\BibitemShut {NoStop}%
\end{thebibliography}
\end{document}